\documentclass[aps,pra,reprint,superscriptaddress,floatfix,nofootinbib]{revtex4-2}

\usepackage{graphicx}
\usepackage{amsmath}
\usepackage{braket}
\usepackage{xcolor}
\usepackage{lipsum}
\usepackage{hyperref}
\usepackage{comment}

\hypersetup{
    colorlinks=true,
    citecolor=teal,
    linkcolor=teal,
    filecolor=teal,      
    urlcolor=teal,
}

\usepackage[nameinlink,capitalise]{cleveref}
\newcommand{\expect}[1]{\langle #1 \rangle}

\usepackage{physics}
\usepackage[utf8]{inputenc}
\usepackage{longtable}
\usepackage{tabularx}

\begin{document}

\title{Quantum quenches in a spin-1 chain with tunable symmetry}

\author{Luis Eduardo Ramos-Sol\'is}
\email[]{luises@estudiantes.fisica.unam.mx}
\affiliation{Instituto de F\'isica, Universidad Nacional Aut\'onoma de M\'exico, Apartado Postal 20-364, 01000 Ciudad de M\'exico, Mexico}
\author{Sayan Choudhury}
\email[]{sayanchoudhury@hri.res.in}
\affiliation{Harish-Chandra Research Institute, Chhatnag Road, Jhunsi,
Prayagraj 211 019, India}
\affiliation{Homi Bhabha National Institute, Training School Complex,
Anushakti Nagar, Mumbai 400 094, India}
\author{Freddy Jackson Poveda-Cuevas}
\email[]{jacksonpc@fisica.unam.mx}
\affiliation{Investigador por M\'exico - Instituto de F\'isica, Universidad Nacional Aut\'onoma de M\'exico, Apartado Postal 20-364, 01000 Ciudad de M\'exico, M\'exico}
\author{Eduardo Ibarra-Garc\'ia-Padilla}
\email[]{eibarragp@g.hmc.edu}
\affiliation{Department of Physics, Harvey Mudd College, 301 Platt Blvd, Claremont, CA 91711, USA}

\date{\today}

\begin{abstract}
In recent years, the dynamics of interacting quantum systems far from equilibrium have attracted significant research interest. Driven by rapid progress in quantum simulators, various non-equilibrium phenomena have now been realized experimentally. In this work, we use the time-evolving block decimation (TEBD) method to investigate the dynamics of an anisotropic spin-1 Heisenberg chain for a wide range of experimentally accessible initial states. By adjusting the parameter $J_q$ that controls the quadrupolar interaction strength, we can tune the system from a non-integrable SU(2) Heisenberg model to an integrable SU(3) Heisenberg model. We examine the local magnetization, entanglement entropy, and spin correlations, and characterize their dependence on $J_q$. We identify a new conserved quantity at the SU(3) symmetric point and provide a theoretical framework to explain our numerical observations in terms of the number of accessible states permitted by this conservation law. Our results provide a route to realize a rich array of non-equilibrium behavior in spin-1 lattice models, which can be engineered in several experimental platforms such as ultracold atoms in optical lattices.
\end{abstract}

\maketitle

\section{Introduction}
\label{sec:Introduction}

The far-from-equilibrium dynamics of interacting quantum systems has been a topic of intense research in recent years~\cite{Nandkishore2015,Altman2018,Abanin2019,Mori2018}. This has been propelled by rapid advances in the development of quantum simulators, thereby leading to the experimental realization of a plethora of non-equilibrium phenomena such as many-body scars~\cite{Chandran2023,Serbyn2021,Moudgalya2022}, Hilbert space fragmentation~\cite{Sala2020,Khemani2020,Moudgalya2022hilbert}, dynamical quantum phase transitions~\cite{Heyl2018}, and time crystals~\cite{Else2020,Sacha2018,Sacha2020,Khemani2019,Zaletel2023}. These theoretical and experimental investigations have led to not only the development of deep insights about the nature of quantum thermalization and quantum chaos~\cite{Borgonovi2016,DAlessio2016} but also the development of quantum technologies~\cite{Moon2026,Li2023}. It is anticipated that non-equilibrium protocols will play a pivotal role in the development of designer quantum materials~\cite{Oka2019,Rudner2020,Weitenberg2021} and quantum sensors~\cite{Montenegro2025,Agarwal2025}.

Concurrently with the developments in exploring non-equilibrium dynamics, another important recent advancement in ultracold atomic physics has been the realization of the SU($N$)-symmetric Fermi-Hubbard model (FHM) with alkaline-earth atoms~\cite{Taie2012,Pagano2014,Zhang2014,Hofrichter2016,Pasqualetti2024}. Intriguingly, in these systems $N$ is tunable and can be as large as 10~\cite{Cazalilla2014,Fallani2023,Gas2026,EIGP_SC_2025}; a new proposal suggests that it may be possible to tune $N$ to even higher values (up to 26) by employing ultracold molecules~\cite{Mukherjee2025}. These experiments have paved the path to realizing and probing new forms of quantum matter with controllable non-Abelian symmetries~\cite{Goban2018,Song2020,Taie2022,Tusi2022,He2025}. In this context, it is worth noting that the quantum phases of SU($N$) lattice models (for $N>2$) have been extensively investigated theoretically~\cite{Papanicolau1988,Honerkamp2004,Xu2008,Gorshkov2010,Manmana2011,Mamaev2022,Nataf2014,Xu2018,Yamamoto2020,Botzung2024,Botzung2024numerical,Zhang2025}. A natural research direction then is to examine the non-equilibrium behavior of these SU($N$)-symmetric systems when $N>2$.

\begin{figure}
    \centering
\includegraphics[width=0.85\linewidth]{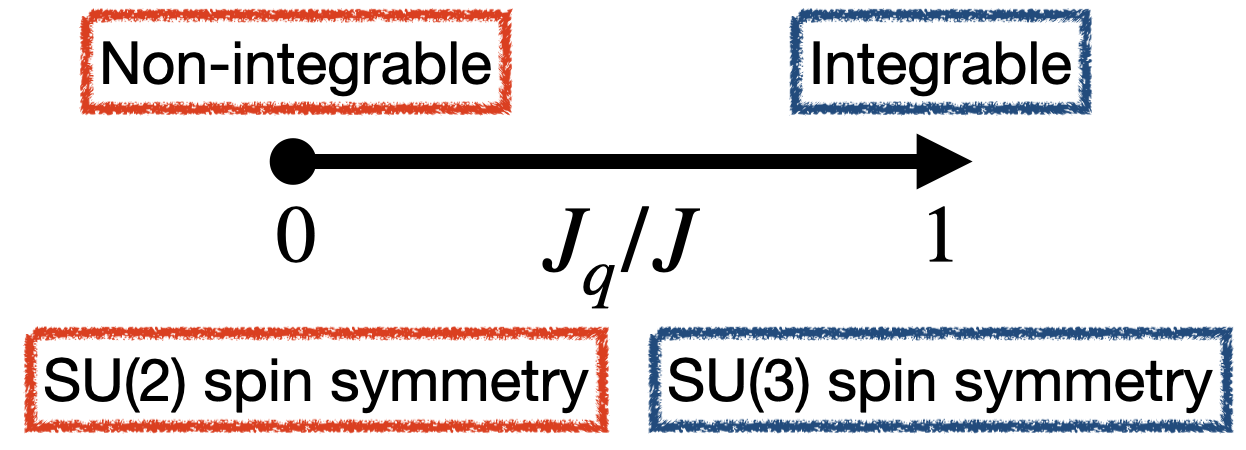}
    \caption{Schematic illustration of the model presented in Eq.~\ref{eqn:anisotropic_H}. This system can be tuned from the non-integrable SU(2) symmetric system to an integrable SU(3) symmetric system by tuning the parameter $J_q/J$ which characterizes the strength of the quadrupolar operator interactions.} 
    \label{fig:integrability}
\end{figure}

In this work, we take an important step in this direction by analyzing the quench dynamics of a spin-1 model which can be tuned from the non-integrable SU(2)-Heisenberg model to the integrable SU(3)-symmetric Heisenberg model by varying a parameter, $J_q/J$. A schematic illustration of this procedure is shown in Fig.~\ref{fig:integrability}. We note that the SU($N$)-Heisenberg model emerges in the strong coupling limit of the SU($N$)-FHM at $1/N$-filling, and it can thus be realized using ultracold alkaline earth atoms in an optical lattice~\cite{EIGP_SC_2025}. We explore the dynamics of this system starting from a wide set of experimentally accessible initial states. These include both eigenstates of the z-magnetization (Fig.~\ref{fig:initial_states}), as well as phantom helix states (Fig.~\ref{fig:initial_states_helix}). We examine the dynamics of the local magnetization, entanglement entropy, fidelity (or survival probability), and various correlation functions, and characterize their dependence on the parameter $J_q/J$. Intriguingly, we demonstrate that a new conserved quantity emerges when $J_q/J = 1$. Consequently, the time evolution of different initial states is governed by the number of accessible states permitted by this conservation law. Notably, both freezing and fidelity revivals occur when the system is initially prepared in nematic states. Interestingly, for the phantom helix initial states, increasing $J_q/J$ leads to faster thermalization, despite the integrability that emerges when $J_q/J = 1$. Our results provide a route to realize a rich array of non-equilibrium behavior in spin-1 lattice models.

This paper is organized as follows. We introduce the model and discuss the initial states, observables, and the numerical techniques employed in this work in Sec.~\ref{sec: Model and Methods}. We present the results of the time-evolution of the system for the z-magnetization initial states, and provide a theoretical framework to explain our numerical observations in Sec.~\ref{sec: Results}. We present results for the time evolution of the system from phantom helix initial states in Sec.~\ref{sec:results_helix}. We conclude with a summary of the results and an outlook for future research in Sec.~\ref{sec:conclusions}.

\section{Model and Methods}
\label{sec: Model and Methods}

\subsection{Hamiltonian}
We study the dynamics of an anisotropic version of the SU(3) Heisenberg Hamiltonian:
\begin{align}\label{eqn:anisotropic_H}
	H &= J_z\sum_{i}S^{z}_i S^{z}_{i+1} + J_{xy}\sum_{i}\left (S^{x}_i S^{x}_{i+1} + S^{y}_i S^{y}_{i+1}\right) \nonumber \\
	&+ J_q\sum_{i}\mathbf{Q}_i\cdot  \mathbf{Q}_{i+1},
\end{align}
where $i=-L/2,\dots,L/2$ labels the sites of a one-dimensional chain with $L$ sites, $\mathbf{S}_i = \left(S^{z}_i,S^{x}_i,S^{y}_i\right)$ are the spin-1 operators at site $i$, and $\mathbf{Q}_i = \left(Q^{x^2-y^2},Q^{z^2}_i,Q^{xy}_i,Q^{yz}_i,Q^{xz}_i\right)$ are the quadrupolar operators at site $i$
\begin{subequations}
    \begin{align}
    Q^{x^2-y^2}_{i} &=  (S^x_i)^2-(S^y_i)^2\\
    Q^{z^2}_i &=  \sqrt{3}(S^z_i)^2-2/\sqrt{3}\\
    Q^{xy}_i &=  S^x_iS^y_i+S^y_iS^x_i\\
    Q^{yz}_i &= S^y_iS^z_i+S^z_iS^y_i \\
    Q^{xz}_i &= S^x_iS^z_i+S^z_iS^x_i
    \end{align}
\end{subequations}

In this study, we consider open boundary conditions (OBC), we set $\hbar = 1$, fix the energy scale $J_{xy}=J=1$, and introduce anisotropic quadrupolar interactions by varying $J_q$ from 0 to 1. 

The Hamiltonian presented in eq.~\eqref{eqn:anisotropic_H} is the Bilinear-Biquadratic model (BB)
\begin{equation}
\label{eqn:Bilinear_Biquadratic}
    H_{BB} = J'\sum_i \left[ \cos \gamma (\mathbf{S}_i \cdot \mathbf{S}_{i+1}) + \sin \gamma (\mathbf{S}_i \cdot \mathbf{S}_{i+1})^2 \right]
\end{equation}
in the limit  $J' = \sqrt{2}J$, $\gamma = \pi/4$ and $J_q=J_z=J$ (see Appendix~\ref{app:equivalence_BB_SU3} for the proof). The BB model is interesting due to the rich phases of matter it displays, such as ferromagnetism, antiferromagnetism, semi-ordered, and nematic phases \cite{Affleck1987, Papanicolau1988, Lauchli2006, Pires2014, Yu2015, Luo2016, Lai2017}. Furthermore, it has recently been proved that it is an integrable model at $\gamma = \pm \pi/4, \pm \pi/2, \pm 3\pi/4$ \cite{Tsvelik1990, Bingtian2022, Park2024}. 

The Hamiltonian considered in this study has two interesting limits: In the limit when $J_q=0$ and $J_z = J_{xy}$, the model reduces to the spin-1 Heisenberg Hamiltonian $H_0 = \sum_i \mathbf{S}_i\cdot  \mathbf{S}_{i+1}$ which is invariant under rotations in spin space described by the SU(2) symmetry algebra. In the limit when $J_q=J_z=J_{xy}=1$, the model corresponds to the SU(3) symmetric version of the Heisenberg Hamiltonian $H_1 = \sum_{i} \mathbf{\Lambda}_i \mathbf{\Lambda}_{i+1}$ where $\mathbf{\Lambda}_i = (\lambda_1,\lambda_2,\dots,\lambda_8)_i$ and $\lambda_j$ are the Gell-Mann matrices \cite{Sasaki1982}. We note that the introduction of the $J_q$ terms adds considerable richness to the spin-1 XXZ model. On the one hand, these terms permit changes in the spin projection such that $\Delta m = \pm 2$, thereby competing against the terms that favor spin anti-alignment ($S_i^zS_j^z$) and processes of the kind $\Delta m = \pm 1$ ($S_i^+S_j^-$). On the other hand, it enables us to tune the underlying symmetry of the model from $SU(2)$ ($J_q=0$) to $SU(3)$ ($J_q=1$). This process concurrently dials the system from a non-integrable point to an integrable one.

The Hamiltonian commutes with the magnetization operator $M = \sum_i S_i^z$ for all values of $J_z$, $J_{xy}$, and $J_q$. This implies that the total magnetization is conserved at all times. Furthermore, when $J_q= J_{xy}$, a new conserved quantity emerges: the \textit{quadratic magnetization} $M^2 = \sum_i (S_i^z)^2$. As we will see in the Results section, the conservation of $M^2$ limits the number of accessible states of the system during its dynamics, thereby preventing thermalization in certain cases. For details on the proofs, see Appendix~\ref{app:conserved_quantities}.

\subsection{Observables}\label{sec:observables}

We are interested in understanding the time dynamics of local and global observables for different initial states. For an arbitrary operator $\mathcal{O}$ at site $i$, its expectation value at time $t$ is computed as $\expect{\mathcal{O}_i} =\bra{\psi(t)}\hat{O}_{i}\ket{\psi(t)}$. 

We compute the local magnetization $\expect{S^{z}_{i}}$, the in-plane $\text{Re}\left(\expect{S^{+}_i S^{-}_{i+1}}\right)$ and out-of-plane $\expect{S_i^zS_{i+1}^z}$ spin-spin correlation functions, and the quadrupolar-quadrupolar correlation functions $\expect{Q^{z^2}_{i}Q^{z^2}_{i+1}}$ and $\expect{Q^{x^2-y^2}_{i}Q^{x^2-y^2}_{i+1}}$. Due to our choice of OBC, we calculate local observables and nearest-neighbor correlation functions at the middle of the chain ($i=0$) to minimize finite-size effects.

We also compute the fidelity
\begin{equation}
    \mathcal{F} = \left|\bra{\psi(0)}\ket{\psi(t)}\right|^2,
\end{equation}
and the half-chain entanglement entropy 
\begin{equation}
    \mathcal{S} = -\sum_{i} \lambda_{i}^2 \ln\left( \lambda^{2}_{i} \right),
\end{equation}
where $\lambda_i$ are the Schmidt coefficients of the Schmidt decomposition of the state $\ket{\psi} = \sum_i \lambda_i \ket{i_L}\ket{i_R}$. Here $\{\ket{i_L}\}, \{\ket{i_R}\}$ form the basis for the left and right subsystems' Hilbert spaces, and $\lambda_i^2$ are the eigenvalues of the reduced density matrices. 

The above observables provide valuable insights into the dynamics of the spin chain. Firstly, the local magnetization describes how the spin at the center of the chain varies over time. Secondly, the half-chain entanglement entropy serves as a quantifier of the number of accessible states the system possesses \cite{DeChiara_2006}. The fidelity (or survival probability) is a measure of the similarity between the time-evolved state and the initial state. This quantity determines whether the system has evolved toward an orthogonal state or has recovered the initial state through revivals associated with integrable systems \cite{Park2025, Rigol2007, Deutsch1991}. Finally, the correlation functions quantify the kind and degree of alignment between neighboring sites along the chain of the center. These alignments are introduced and described in the following sections.

\subsection{Dipolar and quadrupolar states}

To introduce the different spin alignments, we first need to define some terms in the single-site limit. Let the state of the system be $\ket{\psi}$. 

The state is in a magnetic dipolar state if its spin length is maximal, i.e., $|\bra{\psi}\mathbf{S}\ket{\psi}|^2=1$. These correspond to states $\ket{\pm 1}$, which are eigenstates of $S^z$, as well as $\ket{\pm X}$ and $\ket{\pm Y}$ which are eigenstates of the $S^x$ and $S^y$ operators, respectively ($S^{x}\ket{\pm X}=\pm \ket{\pm X}$ and $S^{y}\ket{\pm Y}=\pm \ket{\pm Y}$).

In contrast, quadrupolar states are such that their spin-length is zero, $|\bra{\psi}\mathbf{S}\ket{\psi}|^2=0$~\cite{Penc_2011}. The spin-1 manifold can be spanned in a basis of quadrupolar states. This basis is time-reversal invariant and is defined as:
\begin{subequations}
\label{eqn:quadrupolar_basis}
	\begin{align}
		\ket{x}&=\frac{i}{\sqrt{2}}\left( \ket{1}-\ket{-1} \right),
		\\
		\ket{y}&=\frac{1}{\sqrt{2}}\left( \ket{1}+\ket{-1} \right),
		\\
		\ket{z}&=-i\ket{0}.
	\end{align}
\end{subequations}

Quadrupolar states are anisotropic. This means that the states in the quadrupolar basis fulfill:
\begin{equation}
\label{eqn:anisotropies_fluctuation}
	\left(S^{a}\right)^2\ket{b}=\left(1-\delta^{ab}\right)\ket{b},
\end{equation}
which shows that spin fluctuations in a quadrupolar state occur in a plane perpendicular to an axis known as the \textit{director}. The director of the basis states $\ket{x}$, $\ket{y}$ and $\ket{z}$ are $e_x$, $e_y$ and $e_z$, respectively.

Any spin-1 state can be written in terms of the basis states \eqref{eqn:quadrupolar_basis} as:
\begin{equation}
\label{eqn:general_spin_1}
    \ket{\psi}=\sum_{\alpha=x,y,z}d^{\alpha}\ket{\alpha}.
\end{equation}
Here, $\mathbf{d}$ is a complex vector, written as $\mathbf{d}=\mathbf{u}+i\mathbf{v}$,
where $\mathbf{u}$ and $\mathbf{v}$ are real vectors with elements $d^\alpha = u^\alpha + i v^\alpha$. These two vectors satisfy the normalization condition $|\mathbf{u}|^2+|\mathbf{v}|^2=1,$
and the overall phase of $\ket{\psi}$ can be adjusted so that $\mathbf{u}\cdot\mathbf{v}=0$~\cite{Toth2011}.
Given these conditions, the spin length of an arbitrary spin-1 state is given by:
    $\left\langle\mathbf{S}\right\rangle=2\mathbf{u}\times \mathbf{v}$.
Note that magnetic dipolar states correspond to $|\mathbf{u}|^2=|\mathbf{v}|^2=1/2$, while quadrupolar states correspond to either $|\mathbf{u}|^2=0$ or $|\mathbf{v}|^2=0$. In the case $0<\left\langle\mathbf{S}\right\rangle<1$, the director is defined as the largest-magnitude vector.

\begin{figure}[htbp!]
    \centering
    \includegraphics[width=\linewidth]{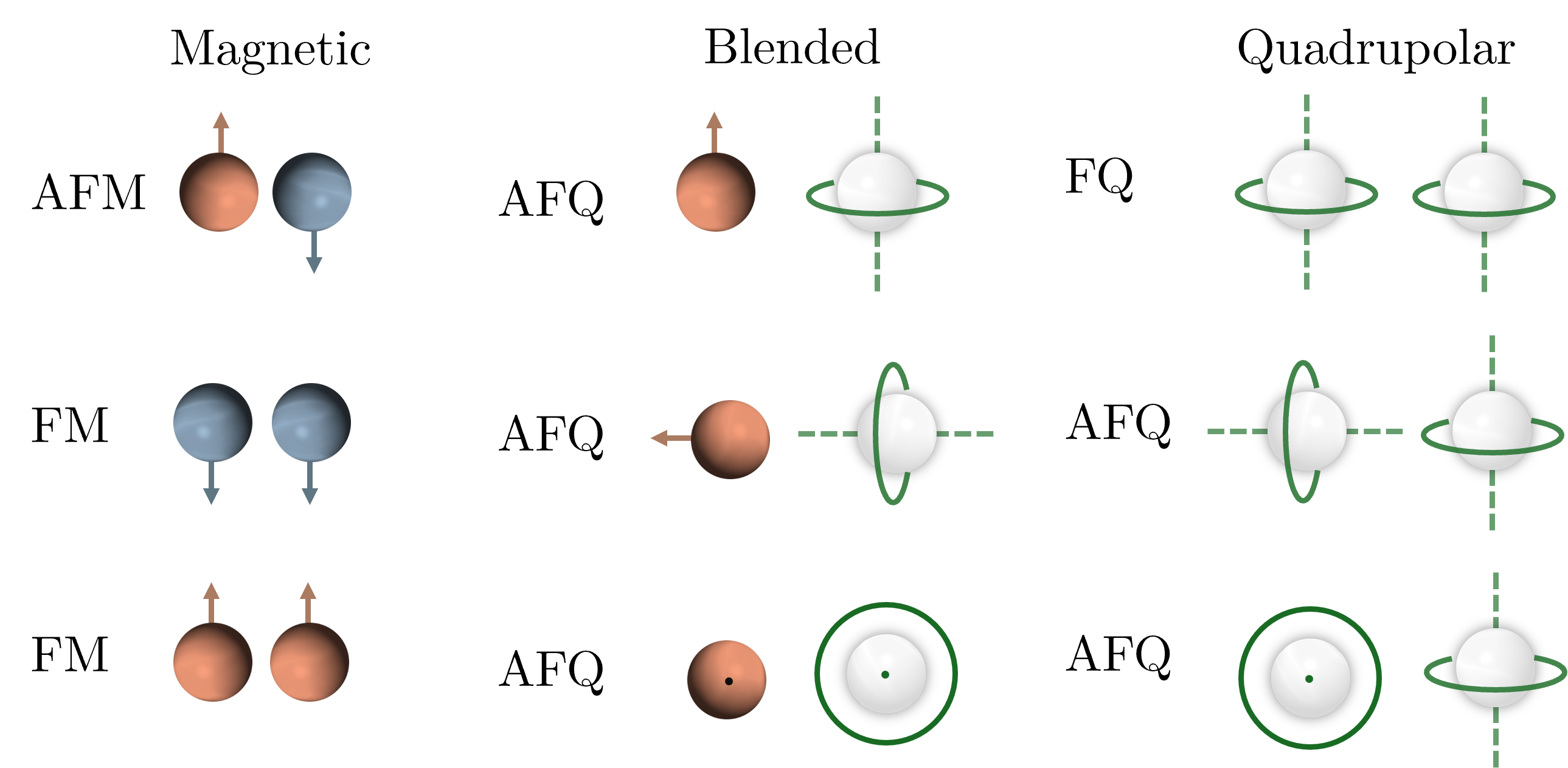}
    \caption{Two-site spin alignments that minimize the energy of Eq.~\eqref{eqn:single_bond_hamiltonian}. Magnetic alignments (left column) feature two dipolar states and can be either FM or AFM. Blended alignments (central column) feature a dipolar and a quadrupolar state, with the dipolar spin vector parallel to the director of the quadrupolar state. This corresponds only to an AFQ alignment. Quadrupolar alignments (right column) feature two quadrupolar states: if both directors are parallel, the alignment is FQ, if they are perpendicular, it is AFQ.}
    \label{fig:alignments}
\end{figure}

\subsection{Two-site framework}
\label{sec:two-site-framework}

\begin{table*}[tbp!]
\caption{\label{tab:magnetic_corr_values}Values of the correlation functions for different types of magnetic, quadrupole-quadrupole, and quadrupole-dipolar alignments. We provide some examples of states that exhibit that type of alignment.
}
\begin{ruledtabular}
\begin{tabular}{cccccc}
Alignment &  Examples & $\langle S^{z}_{0}S^{z}_{1}\rangle$ & $\text{Re}(\langle S^{+}_{0}S^{-}_{0}\rangle)$ & $\langle Q^{z^2}_{0}Q^{z^2}_{1} \rangle$ & $\langle Q^{x^2-y^2}_{0}Q^{x^2-y^2}_{1} \rangle$ \\
\colrule
FM-Z  & $\ket{\pm1}\otimes\ket{\pm1}$ & 1 & 0  & 1/3  & 0 \\
FM-XY &  $\ket{X}\otimes\ket{X}$,\,$\ket{Y}\otimes \ket{Y}$  & 0 & 1  & 1/12 & 1/4 \\
AFM-Z  & $\ket{\pm1}\otimes\ket{\mp1}$ & -1 & 0 & 1/3  & 0  \\
AFM-XY & $\ket{X}\otimes\ket{-X}$,\,$\ket{Y}\otimes \ket{-Y}$ & 0 & -1 & 1/12 & 1/4 \\ 
\hline
FQ-Z & $\ket{0}\otimes\ket{0}$ & 0 & 0 & 4/3 & 0    \\
FQ-XY & $\ket{x}\otimes\ket{x}$,\, $\ket{y}\otimes\ket{y}$  & 0 & 0 & 1/3 & 1    \\
AFQ-XY  & $\ket{x}\otimes\ket{y}$,\, $\ket{y}\otimes\ket{x}$ & 0 & 0 & 1/3 & -1   \\
AFQ-XZ/YZ & $\ket{z}\otimes\ket{y}$,\, $\ket{z}\otimes\ket{x}$  & 0 & 0 & -2/3 & 0   \\
\hline
AFQ-dXY  & $\ket{X}\otimes\ket{x}$,\,$\ket{y}\otimes \ket{Y}$ & 0 & 0 & -1/6 & -1/2 \\
AFQ-dZ  & $\ket{0}\otimes \ket{\pm1}$,\,$\ket{\pm 1}\otimes \ket{0}$ & 0 & 0 & -2/3 & 0 \\
\end{tabular}
\end{ruledtabular}
\end{table*}

To gain physical intuition of the spin alignments in our model, it is useful to first understand its properties in the two-site picture. To that end, we classify the possible alignments of a product state $\ket{\psi}=\ket{\psi_0}\otimes\ket{\psi_1}$ by minimizing the energy of the single bond Hamiltonian
\begin{equation}
\label{eqn:single_bond_hamiltonian}
    H = J\mathbf{S}_0\cdot\mathbf{S}_1 + J_q \mathbf{Q}_0\cdot\mathbf{Q}_1,
\end{equation}
where $\ket{\psi_i}$ is written in the form of equation \eqref{eqn:general_spin_1} with associated vectors $\mathbf{d}_i =\mathbf{u}_{i} + i\mathbf{v}_{i}$. We identify the configurations that minimize the energy of the Hamiltonian under different combinations of $J$ and $J_q$~\cite{Toth2011,Penc_2011}. These are displayed in Fig.~\ref{fig:alignments} and are:

(i) \textit{Ferromagnetic (FM) region ($J<0, J_q =0$)}: Dipolar states on both sites with parallel spin vectors.

(ii) \textit{Antiferromagnetic (AFM) region ($J>0, J_q =0$)}: Dipolar states with anti-parallel spin vectors.

(iii) \textit{Ferroquadrupolar (FQ) region ($J=0,J_q <0$)}: 
Quadrupolar states with parallel directors. This condition follows from expressing the quadrupolar term in terms of $\mathbf{d}_0$ and $\mathbf{d}_1$
\begin{equation}
\label{eqn:identity_quad}
     \langle\mathbf{Q}_{0}\cdot\mathbf{Q}_{1} \rangle= \left|\mathbf{d}_0^{*}\cdot\mathbf{d}_1 \right|^2+ \left|\mathbf{d}_0\cdot\mathbf{d}_1 \right|^2-\frac{2}{3},
\end{equation}
and noticing that the energy is minimized for $\left|\mathbf{d}^{*}_{0}\cdot\mathbf{d}_{1} \right|^2=\left|\mathbf{d}_{0}\cdot\mathbf{d}_{1} \right|^2=1$.

(iv) \textit{Antiferroquadrupolar (AFQ) region ($J=0, J_q >0$)}: A quadrupolar state at one site, and either another quadrupolar state with a perpendicular director, or a spin vector of arbitrary length pointing along the director of the quadrupolar state. This condition follows by noting that eq.~\eqref{eqn:identity_quad} is maximized for $\left|\mathbf{d}^{*}_{0}\cdot\mathbf{d}_{1} \right|^2=\left|\mathbf{d}_{0}\cdot\mathbf{d}_{1} \right|^2=0$.

We calculate $\expect{S^{z}_{0}S^{z}_{1}}$, $\text{Re}\left(\expect{ S^{+}_{0}S^{-}_{1}}\right)$, $\expect{Q^{z^2}_{0}Q^{z^2}_{1}}$ and $\expect{ Q^{x^2-y^2}_{0}Q^{x^2-y^2}_{1}}$ for the spin alignments discussed above. These results are presented in Table~\ref{tab:magnetic_corr_values}. We group the results by the system's symmetry. Since alignments along the $x$ and $y$ axes yield identical correlation values, we denote them as \textit{in plane} (XY) alignments, distinguishing them from \textit{out-of-plane} (Z) alignments along the $z$-axis. This notation is applied to magnetic and quadrupolar orders. A \textit{blended} alignment will be denoted by AFQ-dA, where d indicates the presence of a dipole and A represents the axis along which the spin vector and director are oriented.

\subsection{Initial states}
\label{sec:initial_states}

\begin{figure}[htbp!]
    \centering
    \includegraphics[width=\linewidth]{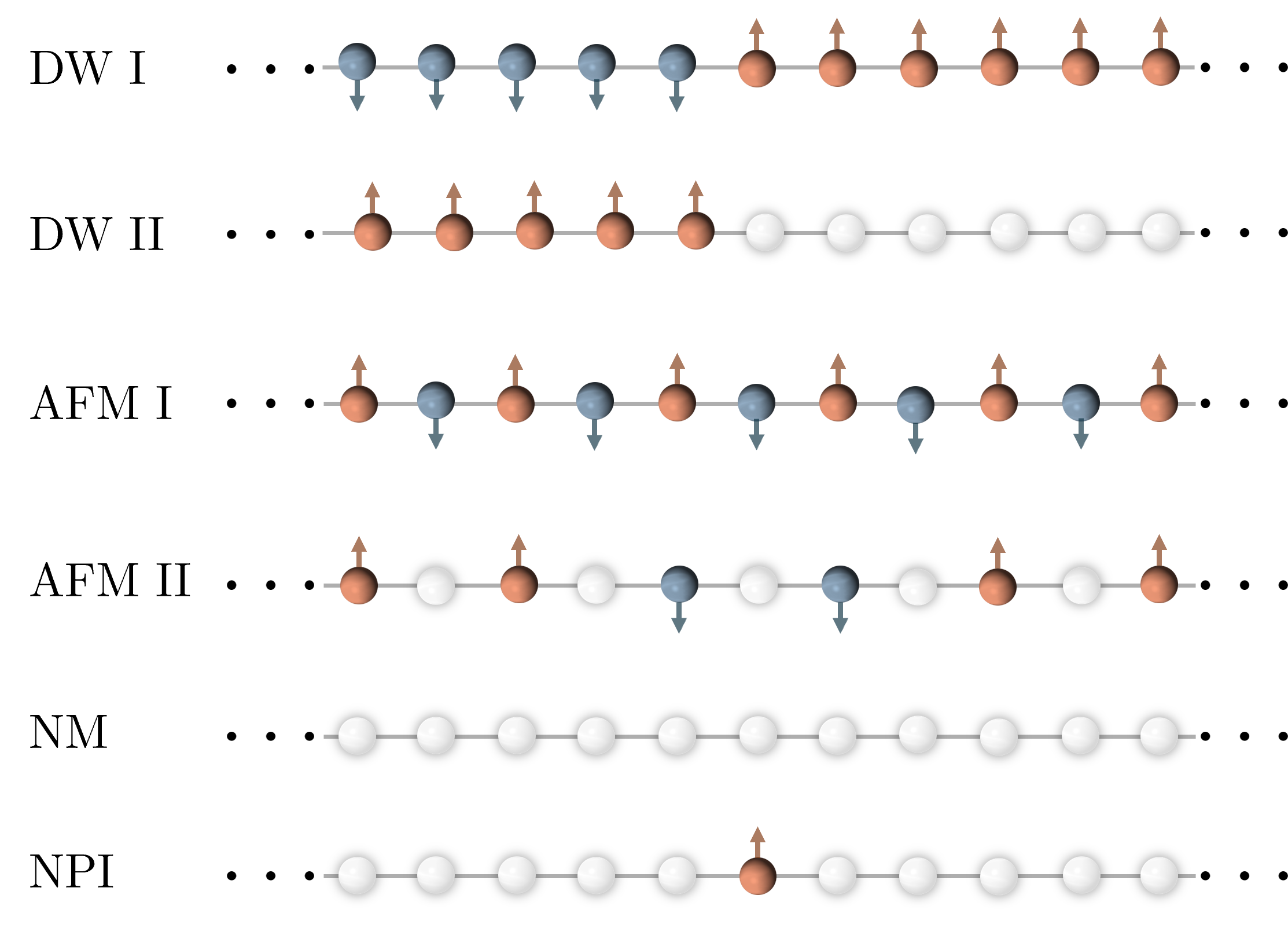}
    \caption{Graphical representation of the z-magnetization initial states we consider in our analysis. A red ball corresponds to $\sigma_i = +1$, a blue ball $\sigma_i = -1$, and a white ball $\sigma_i = 0$. The initial states are the domain-wall I and II (DW I and DW II), the antiferromagnetic I and II (AFM I, AFM II), the nematic (NM), and nematic with polarized impurity (NPI).} 
    \label{fig:initial_states}
\end{figure}

The first set of initial states we consider is depicted in Fig.~\ref{fig:initial_states}. These states are eigenstates of the magnetization operator $M$ and can be represented as product states,
\begin{equation}
\ket{\psi(0)}=\ket{\sigma_{1},\dots,\sigma_{i},\dots,\sigma_{L}},
\end{equation}
where $\sigma_{i} = 0, \pm1$.

\begin{figure*}[tbp!]
    \centering
    \includegraphics[width=0.95\linewidth]{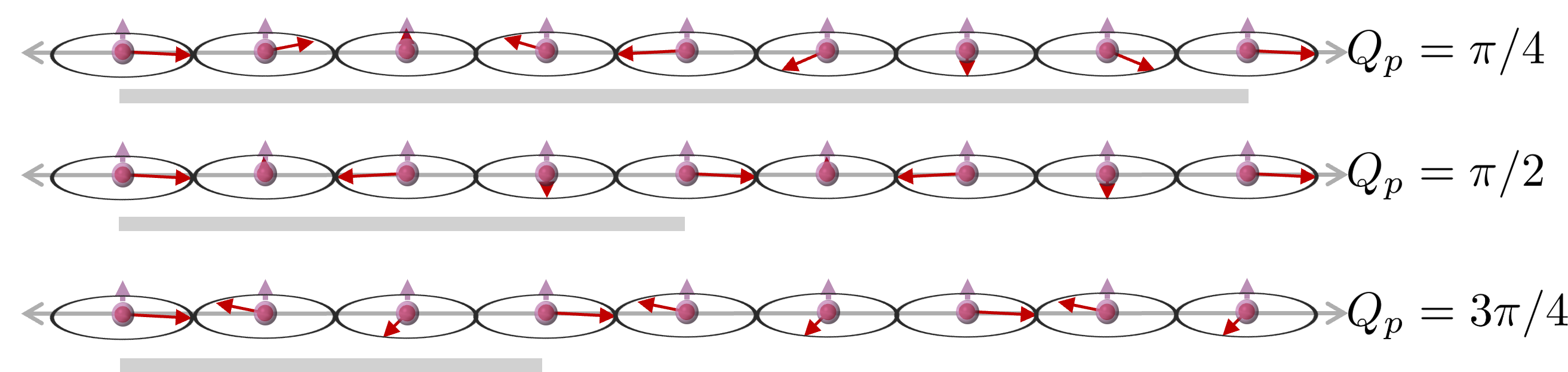}
    \caption{The $xy$-plane spin projections of the phantom helix states represented graphically for fixed $\theta = \pi/2$ and winding angles $Q_p =\pi/4$, $Q_p =\pi/2$ and $Q_p =3\pi /4$. The bar below each chain indicates the period of the helix.} 
    \label{fig:initial_states_helix}
\end{figure*}

We study two domain wall states - DW I and DW II - where a domain wall at the center of the chain separates two regions with distinct magnetic order. The DW I consists of two FM regions with opposite spin projections. The dynamics of the DW I state in the XXZ Heisenberg model for spin-$1/2$ particles have been studied and found to exhibit diffusive behavior~\cite{Misguich2017}. Such prior knowledge makes this state ideal for studying its spin-1 counterpart. In contrast, the DW II has no spin-$1/2$ counterpart, since this state consists of an FM domain and an FQ domain (spin projection 0). This state provides useful insight into how the spin-1 quadrupolar degrees of freedom modify the dynamics. Both states are readily accessible experimentally with ultracold atoms.

The classical N\'{e}el state, labeled AFM I, is the ground state of eq.~\eqref{eqn:anisotropic_H} in the limit $J_q=J_{xy}=0$ and $J>0$. It is a generalization of the spin-$1/2$ Néel state, which exhibits zero-net magnetization and has been shown to preserve traces of the original AFM order in its long-time dynamics~\cite{Pozsgay2013}. The state labeled AFM II corresponds to a spin wave with 8-site periodicity and net magnetization zero. This state was constructed by considering a chain where two consecutive sites are in the same spin projection, and then introducing a spin $\sigma=0$ in between each bond. The AFM II state is interesting because the central sites of the three-site manifolds initialized as $|\pm1,0,\mp1\rangle$ will exhibit zero magnetization throughout the dynamics, as we will show in the Results section. 

Finally, the nematic state (NM) is characterized by zero local z-magnetization $(\sigma^{z}_{i}=0)$ everywhere. Notably, the NM state does not have a spin-$1/2$ counterpart, and the time-evolution of the local magnetization $S_0^z$ is insufficient to understand its dynamics. As shown in Appendix~\ref{app:conserved_quantities}, the NM state is an eigenstate of the Hamiltonian when $J_{xy}=J_{q}$ due to the $M^2$-conservation law. We also consider a special case of an nematic state with a polarized impurity (NPI) in which $\sigma^{z}_{i}=0$ everywhere except at the middle of the chain, where $\sigma^{z}_{0}=1$. As discussed in more detail in Sec.~\ref{sec: Results}, the time-evolution of the NM and NPI states exhibits a strong dependence on $J_q/J$ due to the $M^2$ conservation in the $SU(3)$ limit.

We also study phantom helix states, which have recently been realized with ultracold atoms~\cite{Ketterle2022} in tunable spin-$1/2$ XXZ Heisenberg models. Their importance lies in the fact that these states carry macroscopic momentum yet no energy, and represent a far-from-equilibrium exception to quantum thermalization~\cite{Popkov2021}. Spin helix states are generalized to spin-$1$ as:
\begin{equation}
\label{eqn:helix_state}
    \ket{\psi(Q_p)}=\prod_{i}\left[e^{-\imath Q_p x_i S^z_i}e^{-\imath\theta S^y_i}\ket{1} \right],
\end{equation}
where $S^{z}_i \ket{1}=1\ket{1}$, $\theta$ determines the out-of-plane spin projection $\left\langle S^{z}_{i}\right\rangle=\cos\theta$ and $Q_p$ is the wavevector parameterizing the in-plane winding rate. Fig.~\ref{fig:initial_states_helix} illustrates the plane spin projections of such states. A helix state is known as a phantom helix state if the Hamiltonian satisfies the phantom condition:
\begin{equation}
    J_{z}/J_{xy} = \cos(Q_pa),
\end{equation}
where $a$ is the lattice spacing (set to $1$ in our simulations). Throughout this work, $\theta =\pi/2$ is fixed, and $Q_p$ takes the values $\pi/4$, $\pi/2$ and $3\pi/4$.

Phantom helix states are eigenstates of the $J_q/J=0$ Hamiltonian when periodic boundary conditions (PBC) are considered, and $L$ is a multiple of $\pi/Q_p$. In our simulation, our Hamiltonians exhibit OBC, so these states appear as an \textit{approximate} eigenstate at $J_q/J = 0$. 

\subsection{Numerical methods}

We employ time-evolving block decimation (TEBD) and exact diagonalization (ED) to numerically compute the time evolution of the initial states $\ket{\psi(t)} = e^{-i Ht /\hbar} |\psi(0)\rangle$ and evaluate the expectation value of observables. 

\vspace{-1em}
\subsubsection{Exact diagonalization}

We block-diagonalize eq.~\eqref{eqn:anisotropic_H} in fixed magnetization sectors~\cite{Jung2020} and employ ED to simulate small chains up to $L\sim 12$ sites. ED results were used to benchmark convergence of our TEBD simulations at short times and to diagnose finite-size effects (see Appendix~\ref{app:finite_size_errors} for a discussion). 

\vspace{-1em}
\subsubsection{Time evolving block-decimation}

Tensor network algorithms exploit the entanglement area laws of quantum many-body systems, whose states access only a subspace of the exponentially growing Hilbert space. These algorithms include DMRG (Density Matrix Renormalization Group)~\cite{White1992, White1993} and TEBD~\cite{Vidal2004}.

In this work, we implement the TEBD algorithm to compute the time evolution of the different initial states after a quench. We represent the wavefunction of the initial state as a matrix product state (MPS)
\begin{equation}
\label{eqn:psi_mps}
	\ket{\psi}=\sum_{\boldsymbol{\sigma},\boldsymbol{\alpha}}B^{\sigma_1}_{\alpha_0 \alpha_1 }B^{\sigma_2}_{\alpha_1 \alpha_2} \dots B^{\sigma_L}_{\alpha_{L-1}\alpha_{L}}\ket{\boldsymbol{\sigma}} ,
\end{equation}
where $\boldsymbol{\alpha}=(\alpha_0,\alpha_1,\alpha_2,\dots,\alpha_L)$ are the bond indices of the rank-3 tensors $B^{\sigma_i}_{\alpha_{i-1}\alpha_{i}}$ and $\boldsymbol{\sigma}=(\sigma_0,\sigma_1,\dots,\sigma_L)$ denotes the different configurations of the magnetization basis.

For a fixed system size, TEBD algorithms have two systematic errors: A Trotter error and a bond dimension error. The Trotter error arises from the time evolution operator, in which a Trotter-Suzuki decomposition is employed to time evolve the state by a step size $dt$. In our simulations, $dt = 5\times 10^{-3}$ is chosen as the Trotter step. This value yields small, controlled systematic errors in the simulated times (see Appendix~\ref{app:trotter_explanation}). Smaller Trotter steps incur higher computational cost, severely constraining the time available for our calculations. 

The bond dimension error arises from the Hilbert-space explosion. This means that a maximum bond dimension $\chi$ is defined such that if the bond indices $\alpha>\chi$ or $\beta > \chi$, only the largest $\chi$ singular values in the Schmidt decomposition contribute to the local Hilbert space. Unless otherwise stated, $\chi = 600$ is fixed as the bond dimension in our simulations. This value of $\chi$ yields precise results as compared to larger values of the bond dimension (see Appendix~\ref{app:decimation}). It is worth noting that the decimation error depends solely on the Trotter step for single-magnetization eigenstates and is more pronounced for phantom-helix states.

A detailed discussion about the TEBD algorithm and the procedure to compute various observables is presented in Appendix~\ref{app:tebd_details}.

\section{Results for z-magnetization initial states}
\label{sec: Results}

The results are organized as follows. Sec.~\ref{sec:Results_two-body} shows the exact results for the two-site Hamiltonian for three different relevant alignments defined in Sec.~\ref{sec:two-site-framework}. The two-body problem offers an intuitive understanding of how conservation laws affect the dynamics of our observables, especially correlation functions. This is useful since similar effects are observed in the many-body problem, which are presented in Sec.~\ref{sec:Results_many_body} where we study $L\sim 30$ spin chains. In Sec.~\ref{sec:discussion_z}, we provide a theoretical framework to explain our numerical observations. We set $J_z = J$ in these calculations.

\subsection{Two-body problem}
\label{sec:Results_two-body} 

\begin{figure}[tbp!]
    \centering
    \includegraphics[width=\linewidth]{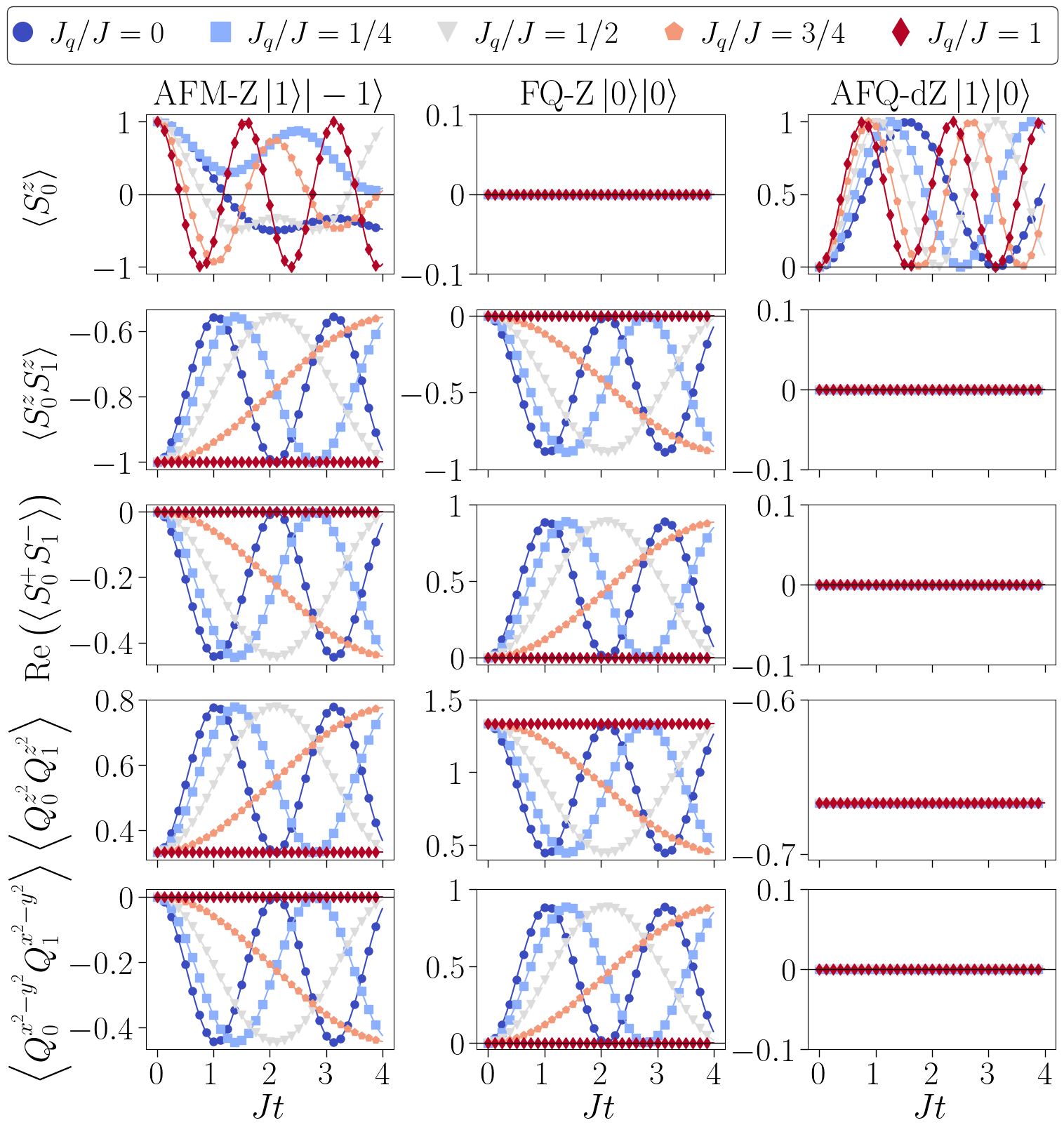}
    \caption{Time evolution of the local magnetization and correlation functions for the single-bond Hamiltonian [Eq.~\eqref{eqn:single_bond_hamiltonian}] for different values of $J_q/J$ for initial states with AFM-Z, FQ-Z, and AFQ-dZ alignments.
    }
    \label{fig:two_site_results}
\end{figure}

Before proceeding to study the full many-body dynamics, it is instructive to examine the two-body quench dynamics. To this end, we study the time evolution of observables of interest for three initial product states in the AFM-Z (first column), FQ-Z (second column), and AFQ-dZ (third column) alignments. Our results are presented in Fig.~\ref{fig:two_site_results}.

The evolution of the out-of-plane local magnetization at the first site $\expect{S^{z}_{0}}$ is displayed in the first row of Fig.~\ref{fig:two_site_results}. We note that $\expect{S_0^z}$ decays from its initial value and oscillates as a function of time for the initial states AFM-Z and AFQ-dZ. The decay is faster for larger values of $J_q/J$, except in the case of $J_q/J=1/4$ for the AFM-Z state. On the contrary, the evolution of the local magnetization for the FQ-Z state remains constant at all times for all values of $J_q/J$. 

In the second and third rows of  Fig.~\ref{fig:two_site_results}, we present the out-of-plane and in-plane spin-spin correlation functions. For the AFM-Z and FQ-Z initial states, we observe that the oscillation frequency decreases as $J_q/J$ increases and completely freezes at $J_q/J=1$. The dynamics of these correlation functions for the AFQ-dZ state are completely frozen at all times for all values of $J_q/J$. Furthermore, the quadrupole-quadrupole correlation functions display a similar behavior as the spin-spin correlation functions, as shown in the fourth and fifth rows of Fig.~\ref{fig:two_site_results}. 

From these two-body systems, we conclude that the correlation functions freeze in two cases: i) when $J_q/J =1$, and ii) when the initial state is in an AFQ-dZ alignment. This freezing is due to the system's conservation laws: when $J_q/J=1$, the model exhibits an $\mathrm{SU}(3)$ symmetry and the quadratic magnetization $M^2$ is a conserved quantity. This implies that the available states in the Hilbert subspace are reduced to those that share the same $M$ and $M^2$ as the initial state. On the other hand, the freezing of correlations in the AFQ-dZ initial state is a consequence of magnetization conservation and of having only two sites. This is because for this system size, only states $\ket{0,1}$ and $\ket{1,0}$ exhibit AFQ-dZ alignment with magnetization $M=1$.

\begin{figure*}[htbp!]
    \centering
    \includegraphics[width=\linewidth]{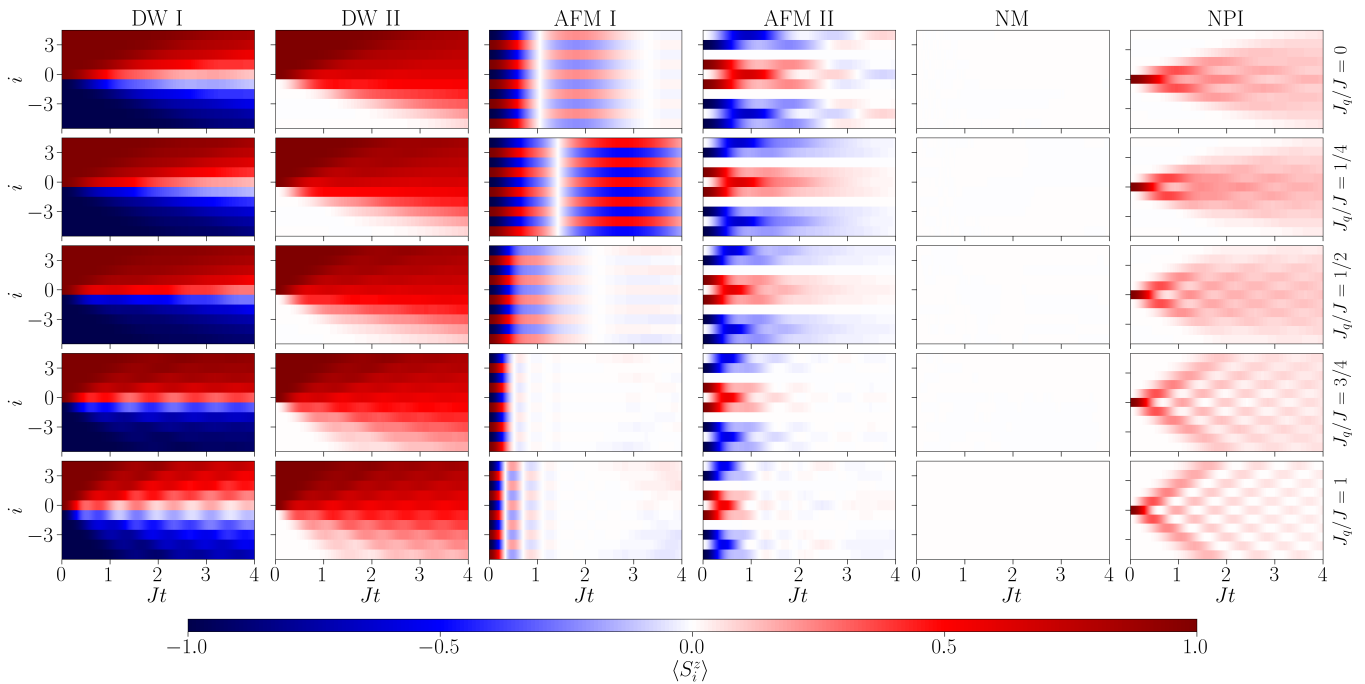}
    \caption{(Color online) The time-evolution of the local magnetization $\langle S_i^z \rangle$ as a function of time $Jt$ at the center-most lattice sites $i$ with $J=J_{z}=J_{xy}$. Results are presented for different values of $J_q/J$ (rows) for different initial states (columns).
    }
    \label{fig:magnetization_heatmaps}
\end{figure*}

\subsection{Many-body problem}
\label{sec:Results_many_body}

We now proceed to examine the many-body dynamics of the spin-chain. To this end, we first study the magnetization transport in real space. Our results presented in Fig.~\ref{fig:magnetization_heatmaps} shows the evolution of the local magnetization $\langle S^{z}_{i}\rangle$ at sites $i\in[-5,5]$ for different initial states. The DW I and DW II states show the merging of two different domains. Within our time window, transport is linear in $t$, with a steeper slope and the presence of clear interference patterns for larger $J_q/J$. On the contrary, the AFM I state does not exhibit a light cone. For a fixed $J_q/J$, all sites cross $\expect{S^{z}_{i}} = 0$ at the same time. The time scale at which this occurs is longer for smaller $J_q/J$ (being largest at $J_q/J =1/4$) and shorter for larger $J_q/J$.
Interestingly, for the AFM II state, the light cone seems to be reduced to regions of three neighboring sites of the form $\ket{\pm 1, 0, \pm 1}$. In other words, it seems like these regions decouple, separated by ``boundary'' sites with frozen local magnetization of $0$. However, the change in sign of $\expect{S_i^z}$ in those three-site manifolds at later times evidences that spin transport between different regions does occur. The reason the local spin projection remains at zero for ``boundary'' sites at all times is that the net spin transport of $1$ and $-1$ into those sites is equal at all times.
A similar mechanism to that in the AFM I state occurs for AFM II: the spin wave structure is robust for small values of $J_q/J$, and the magnetization profile rapidly reaches zero for larger values of $J_q/J$. Finally, we note that the magnetization profile of the NM state remains at zero at every site for all values of $J_q/J$. On the other hand, the NPI state features a well-defined light cone with a slope that is proportional to $J_q/J$. The magnetization in the NPI state spreads across the rest of the chain. 

\begin{figure*}[htbp!]
    \centering
    \includegraphics[width=\linewidth]{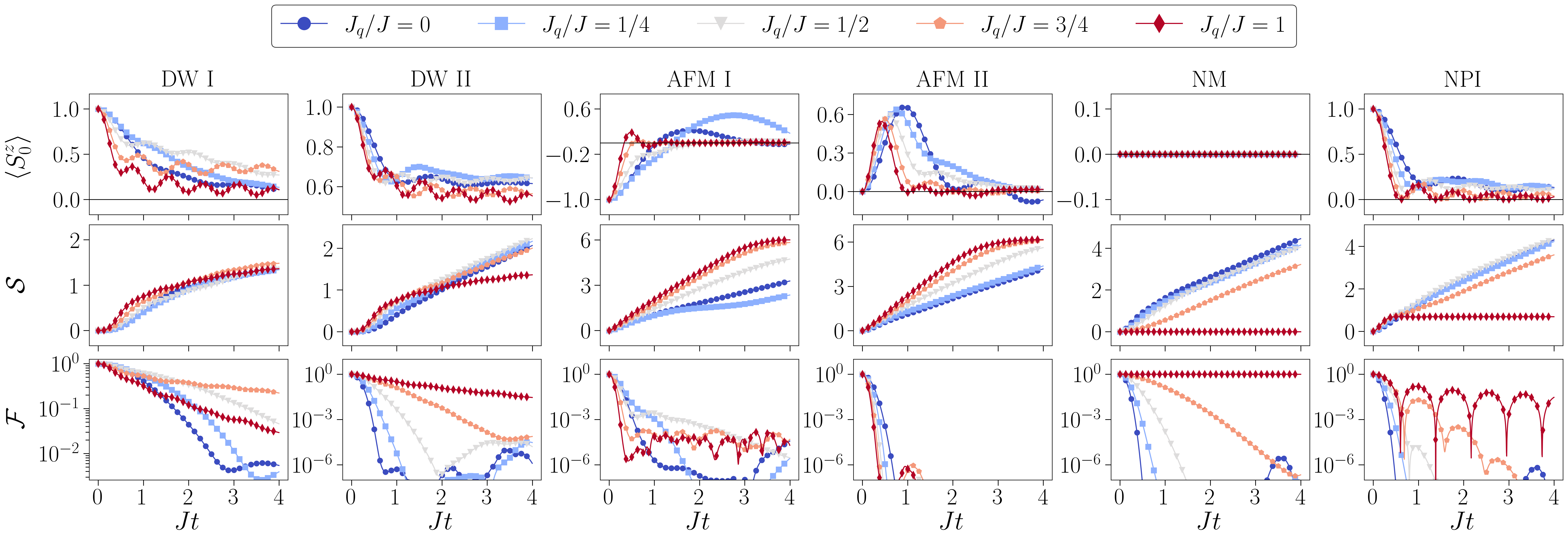}
    \caption{(Color online) Time evolution of the local magnetization $\expect{S^{z}_{0}}$ (upper row), half-chain entanglement entropy $\mathcal{S}$ (middle row), and fidelity $\mathcal{F}$ (lower row) for the initial states depicted in Fig.~\ref{fig:initial_states} for different values of $J_q/J$. Results are presented on lattices of size $L=30$ for DW I, DW II and NM, $L=32$ for AFM I and AFM II, and $L=31$ for NPI.}
    \label{fig:quad_observables}
\end{figure*}

We now focus on the site at the center of the chain to further discuss the $J_q/J$ dependence of the observables. We start with the local magnetization, which is presented in the first row of Fig.~\ref{fig:quad_observables}. For most initial states, the decay of the initial local magnetization is faster for larger $J_q/J$. Such faster decay can be accounted for by the stronger quantum fluctuations provided by the quadrupolar terms as $J_q$ is turned on. More interestingly, the dynamics are highly dependent on the initial state. While the DW I, DW II, and NPI states present oscillations around their asymptotic value (which occur at higher frequencies for larger $J_q/J$), the AFM I and AFM II states exhibit a large-amplitude oscillation at the beginning, after which the curve approaches zero. Finally, the local magnetization of the ZP state remains constant for all values of $J_q$. Quantum fluctuations in this state only occur in the $xy$ plane.

The evolution of the entanglement entropy between two halves of the chain, shown in the second row of Fig.~\ref{fig:quad_observables}, demonstrates how the conservation laws at the SU(3) symmetric point influence the dynamics. For $J_q/J <1$, the initial states studied display the common behavior observed in one-dimensional systems where for a subsystem of size $\ell$, $\mathcal{S}$ grows linearly in time up to a time $t = \ell/2v$, after which they saturate at a value proportional to the subsystem size $\ell$ and the bond dimension $\ln(\chi)$~\cite{Calabrese2005,DeChiara_2006,Fagotti2008,Vincenzo2017}. In the linear regime, the slope is inversely proportional to the velocity $v$ of the elementary excitations~\cite{Calabrese2005}. In the case of the DW I, DW II, AFM I, and AFM II states, the slope of $\mathcal{S}$ grows as $J_q/J$ increases. Moreover, the slow growth of $\mathcal{S}$ at short times $Jt\lesssim0.5$ for DW I and DW II can be attributed to the fact that initially the only region of space where elementary excitations can occur is at the middle of the chain. In contrast, for the NM and NPI states, the slope decreases as $J_q/J$ increases. These results illustrate how the velocity of the quasiparticles depends on $J_q/J$ and the initial state.

The $J_q/J=1$ point merits further discussion. In this limit, the system exhibits an enhanced symmetry, and $M^2$ is conserved. This conservation law restricts the number of accessible states to the system and hinders thermalization. In the NPI state, $\mathcal{S}$ saturates rapidly, at $Jt\approx0.5$, while the NM state is an eigenstate of the Hamiltonian and therefore its time evolution is frozen. Additionally, it is worth noting that while the DW II state shows a fast linear growth of $\mathcal{S}$ at short times ($Jt \lesssim 0.5$), it then exhibits a smaller slope and points to saturating at a smaller value of $\mathcal{S}\sim1.5$.

Finally, while these arguments provide a general understanding of $\mathcal{S}$ with time, the behavior of the entanglement entropy at $J_q/J = 1/4$ remains unexplained: Why does $\mathcal{S}$ grow at almost the same rate as in the $J_q/J =0$ case? And why, for the AFM I state, does it stay below the $J_q/J = 0$ line and then exhibit non-monotonic behavior with a plateau at $Jt \sim 2$? A simple theory that captures these more refined features and $J_q/J$ dependencies could provide useful insights into the physics of the anisotropic Heisenberg model, and our data will serve as an ideal test for any candidate theory.

The fidelity dynamics are presented in the third row of Fig.~\ref{fig:quad_observables}. While $\mathcal{F}$ decays slowly for the DW I and DW II states, the system quickly reaches orthogonality for the AFM I and AFM II states. Except for the DW I state, the behavior of the fidelity with $J_q/J$ is monotonic: For AFM I and AFM II, $\mathcal{F}$ decays faster for larger $J_q/J$. In contrast, for DW II, NM, and NPI, $\mathcal{F}$ decays faster for smaller $J_q/J$. Finally, at the $J_q/J=1$ point we notice that: (1) for DW I and DW II, at $J_q/J=1$, the fidelity displays an exponential decay, as evidenced by the linear trend in the linear-log plot, (2) for the NM state, it remains constant, numerically validating this is an eigenstate of eq.~\eqref{eqn:anisotropic_H} in this limit, and (3) for the NPI state the system exhibits periodic \textit{revivals} with very slow decay. We note that periodic oscillations of physical observables have also been observed in other integrable models, such as the spin-1/2 Heisenberg model, where these oscillations originate from dynamical symmetries~\cite{medenjak2020isolated}.

\begin{figure*}[htbp!]
    \centering
    \includegraphics[width=\linewidth]{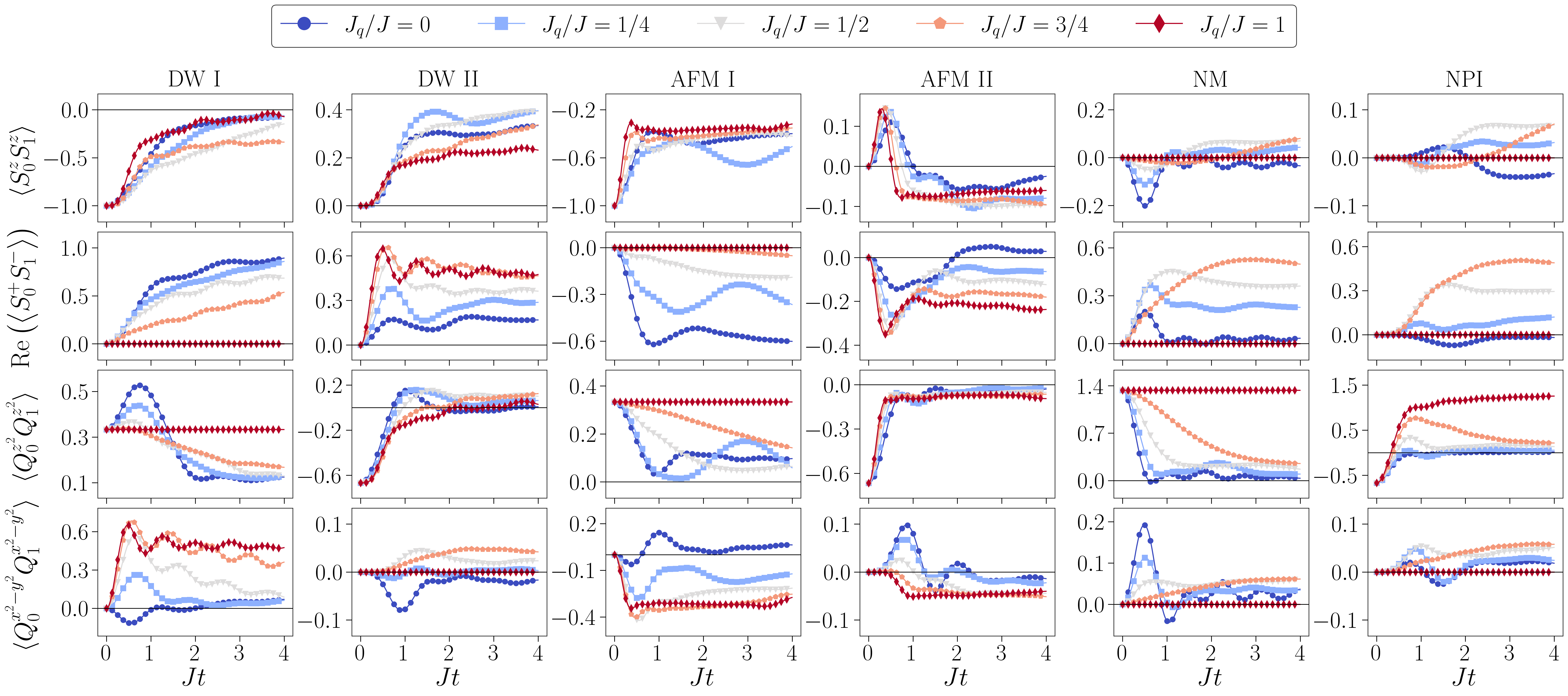}
    \caption{(Color online) Time evolution of the out-of-plane spin correlation $\expect{S^{z}_{0}S^{z}_{1}}$ (first row), in-plane spin correlation $\text{Re}(\expect{S^{+}_{0}S^{-}_{1}})$ (second row), out-of-plane quadrupole correlation $\expect{Q^{z^2}_{0}Q^{z^2}_{1}}$ (third row), and in-plane quadrupole correlation $\expect{Q^{x^2-y^2}_{0}Q^{x^2-y^2}_{1}}$ for the same initial states, values of $J_q/J$, and system sizes as those used in Fig~\ref{fig:quad_observables}.
    }
    \label{fig:quad_correlations}
\end{figure*}

We now proceed to investigate the time-evolution of various correlation functions. The out-of-plane spin-spin correlation dynamics $\expect{S^{z}_{0}S^{z}_{1}}$ are presented in the first row of Fig.~\ref{fig:quad_correlations}. We begin by analyzing the DW I and AFM I states, which start with an AFM-Z alignment at the center of the chain and therefore $\expect{S_0^zS_1^z} = -1$ at $t=0$. For these two states, the magnitude of $\expect{S^{z}_{0}S^{z}_{1}}$ decays in time and finally settles at a negative value. In the case of the DW I state the asymptotic value depends on $J_q/J$ in a non-monotonic way: First, at $J_q/J=0$ it approaches zero. Then, as $J_q/J$ increases, the magnitude of the correlation grows, reaching $-1/3$ at $J_q/J = 0.75$. Finally, since $J_q/J=1$, it returns to 0. On the other hand, for the AFM I state, the asymptotic value is close to -1/3 for all values of $J_q/J$, suggesting that the N\'eel state preserves traces of its original antiferromagnetic order in the long time limit evolution, similarly to what has been shown for spin-$1/2$ systems~\cite{Pozsgay2013}.

We now focus on the DW II, AFM II, NM, and NPI states, which start with an AFQ-dZ alignment at the center of the chain (a dipole neighboring a quadrupole oriented in the $z$-axis) and thus $\expect{S^{z}_{0}S^{z}_{1}} = 0$ at $t=0$. First, for the DW II state, the correlation evolves towards a positive value, consistent with the FM-Z alignment of the first half of the chain. In this case, $\expect{S^{z}_{0}S^{z}_{1}}$ exhibits an initial growth that seems independent of $J_q/J$, after which it keeps growing at a slower rate that depends non-monotonically on $J_q/J$. Second, for the AFM II state, $\expect{S^{z}_{0}S^{z}_{1}}$ exhibits a rapid transient in which the correlation reaches a maximum value before $Jt \sim 0.5$. The time scale for this transient is shorter as $J_q/J$ increases. After this transient, $\expect{S^{z}_{0}S^{z}_{1}}$ quickly evolves towards an small asymptotic negative value. Third, for the NM and NPI states, which are composed of mostly single-site quadrupolar states, the magnitude of $\expect{S^{z}_{0}S^{z}_{1}}$ is very small at all times for $J_q/J<1$. When $J_q/J=1$, the correlation freezes at its initial value.

The time evolution results for the correlation function $\text{Re}(\expect{S^{+}_{0}S^{-}_{1}})$ are shown in the second row of Fig.~\ref{fig:quad_correlations}. Since all the initial states feature dipoles/quadrupoles with spin vector/director parallel to the $z$-axis, the initial value of this correlation is zero. 
As time evolves, for the DW I and AFM I states, the magnitude of $\text{Re}(\expect{S^{+}_{0}S^{-}_{1}})$ is larger for smaller $J_q/J$, being largest at $J_q/J =0$. Conversely, for $J_q/J<1$ for the NM and NPI states, $\text{Re}(\expect{S^{+}_{0}S^{-}_{1}})$ is larger for larger $J_q/J$, reaching its largest value at $J_q /J=3/4$. In the case of the NPI state, the in-plane correlation function attains a positive value after the quench for $J_q/J>0$, but for $J_q/J=0$ it is first negative and then tends to zero.
Finally, we observe that at $J_q/J =1$, this correlation function freezes at its initial value for the DW I, AFM I, NM, and NPI states. Intriguingly, for the DW II and AFM II states, the magnitude of the correlation function is larger as $J_q/J$ increases, attaining its largest value at the SU(3) symmetric point at most times.

The dynamics of the out-of-plane quadrupolar-quadrupolar correlation function $\expect{ Q^{z^2}_{0}Q^{z^2}_{1}}$ is presented in the third row of Fig.~\ref{fig:quad_correlations}. When the initial alignment involves two dipolar states with spin vector oriented parallel to the $z$-axis, such as in the DW I and AFM I states, the initial value of the correlation is $1/3$ (see Table~\ref{tab:magnetic_corr_values}). For these two initial states, when $J_q/J < 1$, the correlation function evolves towards a value $<1/3$, while for $J_q/J=1$ the correlation freezes at $1/3$. When the alignment is AFQ-dZ, such as in the DW II, AFM II, and NPI initial states, this correlation function is minimized and $\expect{ Q^{z^2}_{0}Q^{z^2}_{1}}=-2/3$. For the  DW II and AFM II states, $\expect{ Q^{z^2}_{0}Q^{z^2}_{1}}$ evolves to a value near zero, exhibiting similar dynamics independent of $J_q/J$. However, for the NPI state, the correlation at the SU(3) point shows a contrasting behavior to other $J_q/J$ values: For $J_q/J<1$, $\expect{Q^{z^2}_{0}Q^{z^2}_{1}}$ initially grows and then approaches zero. On the other hand, for $J_q/J=1$ the correlation grows and then saturates near $4/3$, approaching the maximum value of the correlation function for FQ-Z alignment. 
Finally, the NM state initially exhibits FQ-Z alignments in all of its bonds and $\expect{ Q^{z^2}_{0}Q^{z^2}_{1}}=4/3$ at $t=0$. As time passes, the correlation function decays faster as $J_q/J$ decreases, evolving towards zero. For $J_q/J=1$, the correlation remains constant, consistent with the finding that the NM state is an eigenstate of the Hamiltonian in this limit.

The evolution of the in-plane quadrupole-quadrupole correlation function $\expect{Q^{x^2-y^2}_{0}Q^{x^2-y^2}_{0}}$ is presented in the last row of Fig.~\ref{fig:quad_correlations}. Similarly to $\text{Re}(\expect{S^{+}_{0}S^{-}_{1}})$, this correlation function is zero at $t=0$ for all the initial states considered. As time evolves, we observe that for the DW I and AFM I states, the magnitude of the correlation function is larger for larger $J_q/J$. Interestingly, at short times we notice that for $J_q/J=0$, $\expect{Q^{x^2-y^2}_{0}Q^{x^2-y^2}_{0}}$ has an opposite sign than for $J_q/J>0$. This opposite sign persists for the AFM I state at all times. At $J_q/J=1$, $\expect{Q^{x^2-y^2}_{0}Q^{x^2-y^2}_{0}}$ freezes for the DW II, NM, and NPI initial states. What is unexpected is that for the AFM I and AFM II, the correlation function exhibits an almost constant value during $Jt \in [1,3.5]$.

\subsection{Combinatorial analysis of results}
\label{sec:discussion_z}

In the previous sub-section, we have presented numerical results on the quench dynamics in the spin chain.  We now attempt to gain a deeper understanding of the dynamics by characterizing the initial states in terms of the number of accessible states and the relative frequency of out-of-plane alignments in the Hilbert subspace allowed by the initial magnetization $M$, and in the limit of $J_q/J=1$, the quadratic magnetization $M^2$ too.

\vspace{-1em}
\subsubsection{Number of accessible eigenstates}
\label{sec:number_of_states}

\begin{figure}[tbp!]
    \centering
    \includegraphics[width=\linewidth]{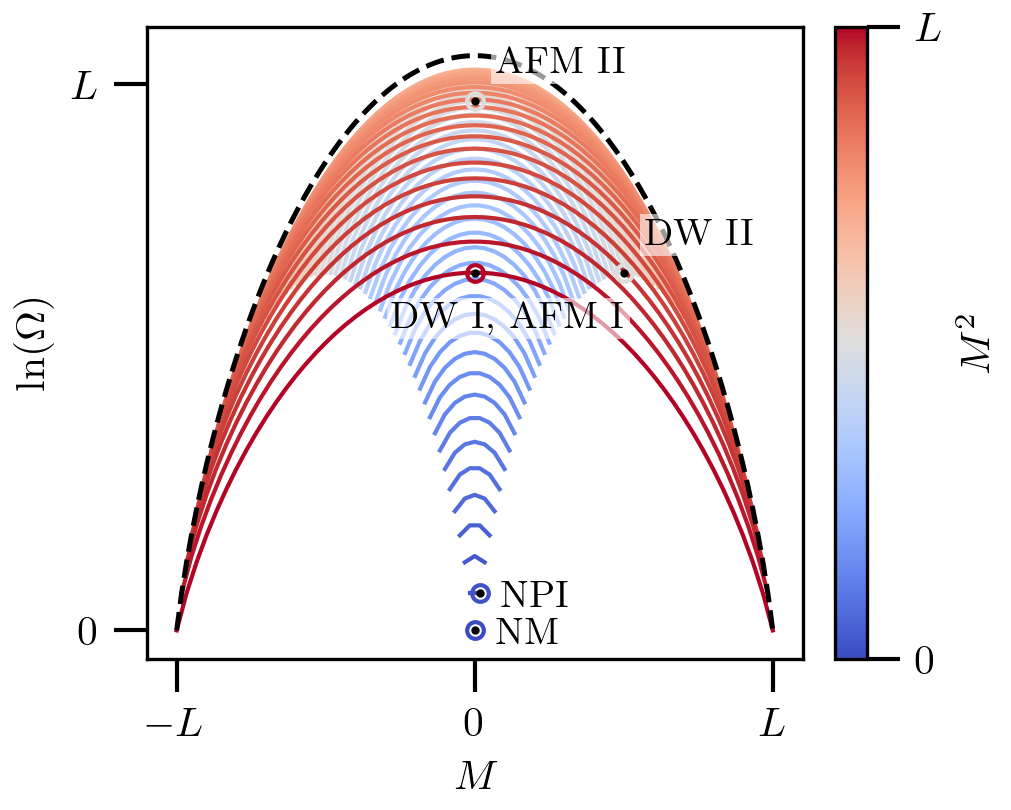}
    \caption{(Color online) Logarithm of the number of available states $\ln(\Omega)$ as a function of magnetization $M$. The solid curves represent different values of the quadratic magnetization $M^2$ with the color scale ranging from blue ($M^2 = 0$) to red ($M^2 = L$). The dashed black line indicates the value of $\ln(\Omega)$ for $J_q\neq J$. Markers correspond to our initial states, characterized by their values of $M$ and $M^2$.}
    \label{fig:mag_blocks_entropy}
\end{figure}

Consider an initial state with a fixed magnetization $M$ and quadratic magnetization $M^2$. Let $\ell_{1}$, $\ell_{-1}$ and $\ell_{0}$ be the number of sites with $\sigma = 1,-1,0$ respectively. Note that $\ell_{1}+\ell_{0}+\ell_{-1}=L$, and that $\ell_{1}-\ell_{-1}=M$. When $J_q/J <1$ only the total magnetization is conserved, and the number of available states can be obtained by summing over the allowed configurations of $k=\ell_{-1}$:
\begin{equation}
\label{eqn:N_states_magnetization_block}
    \Omega(L,M) = \sum_{k=k_{\min}}^{k_{\max}}\frac{L!}{k!(L+M-2k)!(k-M)!}
\end{equation}
where $k_{\min}=\max(M,0)$ and $k_{\max}=\lfloor{\frac{1}{2}\left( L + M\right) }\rfloor$ (where $\lfloor \ldots \rfloor$ denotes the floor operation). We present the 
logarithm of eq.~\eqref{eqn:N_states_magnetization_block} for fixed $L$ as a function of $M$ as the black-dashed line in Fig.~\ref{fig:mag_blocks_entropy}. This curve reaches its global maximum at $M=0$.

When $J_q/J=1$, the quadratic magnetization $M^2$ is a conserved quantity. When this occurs, the variable $k$ in eq.~\eqref{eqn:N_states_magnetization_block} acquires a single available value $k^*=\frac{1}{2}\left(M + M^2\right)$, thus reducing the number of accessible states:
\begin{equation}
\label{eqn:N_states_mag_and_quad_block}
    \Omega(L,M,M^2) = \frac{L!}{k^*! (L+M-2k^*)!(k^*-M)!  }.
\end{equation}
For a combination of $M$ and $M^2$ to be a physically viable solution, the condition for $k^*$ implies that $M + M^2$ must be an even integer. Since $M+M^2=2\ell_{1}$, all single-magnetization states satisfy this criterion. 

We present the logarithm of eq.~\eqref{eqn:N_states_mag_and_quad_block} in Fig.~\ref{fig:mag_blocks_entropy} for fixed $L$ for different values of $M^2$. Similarly to the results for $J_q/J < 1$, the number of available states is maximal at $M=0$ for a fixed $M^2$. The dependence on the quadratic magnetization is non-monotonic, and its global maximum occurs at $M^2 = 2L/3$, which can be proven using Stirling's formula in the limit of $L\gg 1$.

In Fig.~\ref{fig:mag_blocks_entropy}, we also locate the initial states with markers. The NM and NPI are located in the regions with the lowest density of accessible states. The NM is the only accessible state in its Hilbert subspace, thus becoming an eigenstate and showing no dynamics in the $J_q/J = 1$ limit. The NPI state exhibits fidelity revivals precisely because of its restricted Hilbert subspace. On the contrary, the AFM II has the largest number of accessible states, and its dynamics displays a rapid growth of the entanglement entropy and a fast decay of its fidelity for all values of $J_q/J$.

\vspace{-1em}
\subsubsection{Relative frequency of out-of-plane alignments}
\label{sec:relative_frequency_alignments}

\begin{figure}[tbp!]
    \centering
    \includegraphics[width=\linewidth]{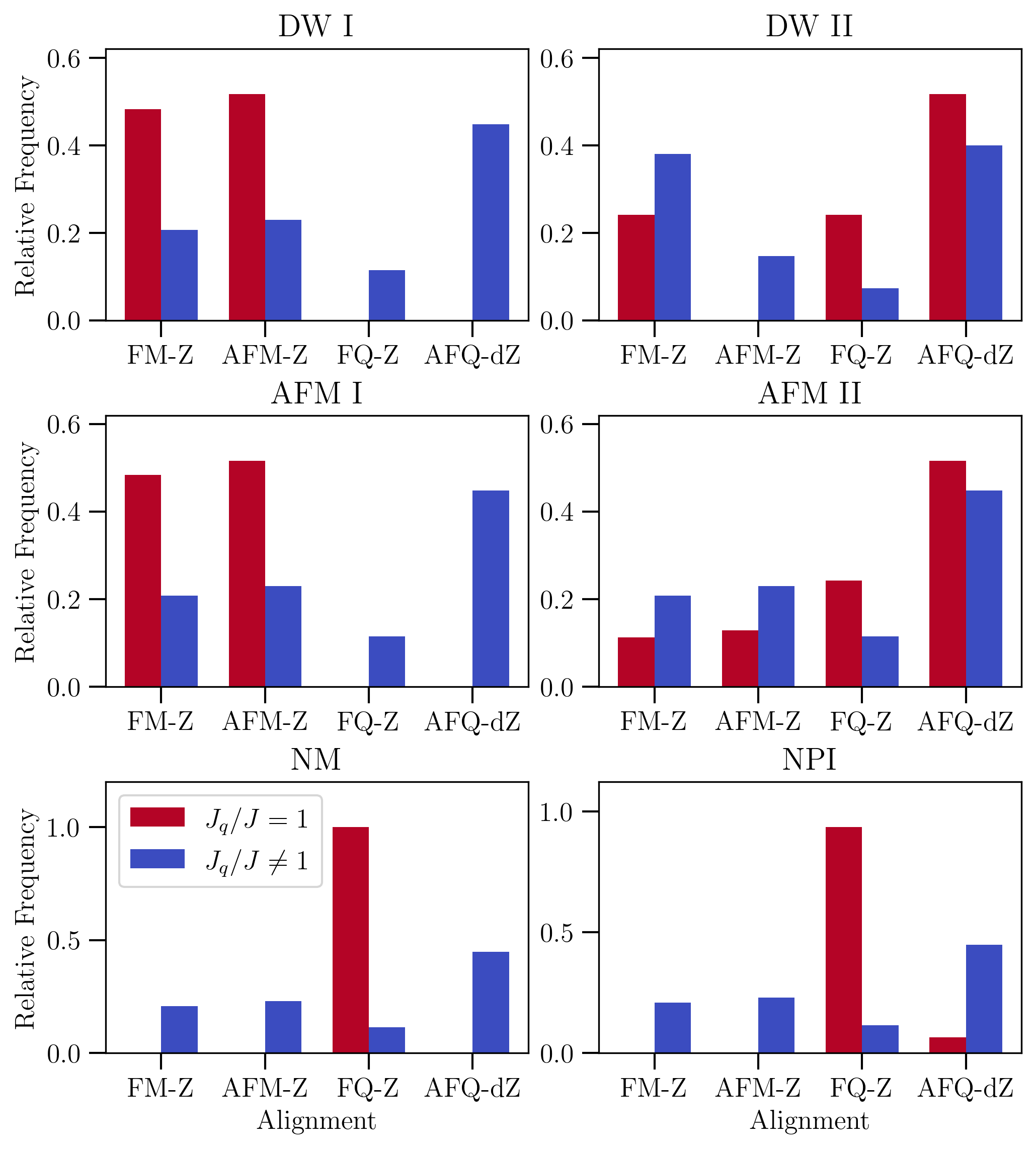}
    \caption{(Color online) Relative frequency of local bond alignments in the Hilbert subspace for different symmetry sectors. The histograms compare the available state space density for $J_q/J = 1$ (red bars) against $J_q/J \neq 1$ (blue bars). The alignments are categorized as FM-Z, AFM-Z, FQ-Z, and AFQ-dZ. Panels correspond to the subspaces of the initial states DW I, DW II, AFM I, AFM II, NM, and NPI.
    }
    \label{fig:alignment_distributions}
\end{figure}

A similar combinatorial argument explains the freezing of certain correlation functions in the $J_q/J = 1$ limit. First, we isolate the center bond, denoted as $\ket{\sigma_0,\sigma_1}$, and define a complementary chain of $\tilde L=L-2$ sites. This reduced chain has a fixed magnetization of $\tilde M = M-M^{\text{bond}}$ and quadratic magnetization $\tilde M^2 = M^2-(M^2)^{\text{bond}}$, where the bond contributions $M^{\text{bond}}=\sigma_{0}+\sigma_{1}$ and $(M^2)^{\text{bond}}=\sigma^2_{0}+\sigma^2_{1}$ depend on the local alignment at the center bond (see Sec.~\ref{sec:two-site-framework} and Fig.~\ref{fig:alignments}). Then, we compute the number of accessible states $\tilde \Omega$ for the complementary chain. This calculation yields the number of states with configuration $\ket{\sigma_0,\sigma_1}$ at the center and fixed magnetization $M$ (and $M^2$ in the $J_q/J=1$ limit). 

The center bond can take one the possible configurations $\ket{\pm1,\pm 1}$, $\ket{\pm 1, \mp 1}$, $\ket{0,0}$, $\ket{\pm1,0}$ and $\ket{0,\pm1}$, which correspond to the FM-Z, AFM-Z, FQ-Z and AFQ-Z alignments. The relative frequency of each of these alignments at the center of the chain is determined by the ratio $\tilde\Omega/\Omega$, where $\Omega$ represents the total number of states in the full chain. These results are presented in Fig.~\ref{fig:alignment_distributions}. We observe that when $J_q/J<1$, all of the four possible alignments are allowed. However, at the $\mathrm{SU}(3)$ point ($J_q/J=1$), the available alignments are constrained depending on the value of $M$ and $M^2$ of the initial state. These restrictions on the Hilbert space have the following consequences for our correlation functions:

\textit{Out-of-plane spin correlation $\expect{S^{z}_{0}S^{z}_{1}}$}: This function takes the value $+1$ when the alignment is FM-Z, $-1$ when its AFM-Z, and $0$ if the alignment is either FQ-Z or AFQ-dZ (see Table~\ref{tab:magnetic_corr_values}). Thus, if the Hilbert subspace exhibits only FQ-Z or AFQ-Z alignments, this correlation freezes at $0$. This is the case of the NM and NPI states, for which $\expect{S^{z}_{0}S^{z}_{1}}=0$ at all times.

\textit{In-plane spin correlation $\text{Re}(\expect{S^{+}_{0}S^{-}_{1}})$}: The operator $S^{+}_{0}S^{-}_{0}$ flips alignments in the following way: AFM-Z into FQ-Z ($S^{+}_{0}S^{-}_{0}\ket{-1,+1}=2\ket{0,0}$), and AFQ-dZ into AFQ-dZ ($S^{+}_{0}S^{-}_{0}\ket{0,+1}=2\ket{+1,0}$). Therefore, this correlation function will be non-zero when: i) the relative frequency of an AFQ-dZ alignment is finite, or ii) when the relative frequencies of both AFM-Z and FQ-Z alignments are finite. For example, the DW II state meets criterion i) and the AFM II state meets both criteria. Thus $\text{Re}(\expect{S^{+}_{0}S^{-}_{1}})$ is not frozen for these initial states. On the contrary, non of the other initial states meet i) nor ii), and therefore $\text{Re}(\expect{S^{+}_{0}S^{-}_{1}})$ remains frozen at 0.

\textit{Out-of-plane quadrupolar correlation $\expect{Q^{z^2}_0Q^{z^2}_1}$}: This function freezes at $+1/3$ if there is only FM-Z or AFM-Z alignments accessible (this is the case for the DW I and AFM I states). Similarly, if the only available alignment is FQ-Z, this function will freeze at $+4/3$ (this is the case for the NM state), and if the only available alignment is AFQ-dZ, it will freeze at $-2/3$. We note that the NPI state has a higher relative frequency of FQ-Z alignment than AFQ-dZ relative frequency. This explains why its $\expect{Q^{z^2}_0Q^{z^2}_1}$ evolves from $-2/3$ (initial AFQ-dZ alignment) towards $4/3$.

\textit{In-plane quadrupolar correlation $\expect{Q^{x^2-y^2}_0Q^{x^2-y^2}_1}$}: Because $Q^{x^2-y^2}_{0}Q^{x^2-y^2}_{1}\ket{0,\pm 1}=Q^{x^2-y^2}_{0}Q^{x^2-y^2}_{1}\ket{\pm 1, 0}=Q^{x^2-y^2}_{0}Q^{x^2-y^2}_{1}\ket{0,0}=0$ and $Q^{x^2-y^2}_{0}Q^{x^2-y^2}_{1}\ket{\pm 1,\pm 1}=\ket{\mp 1, \mp 1}$ this correlation function will be non-zero when the relative frequency of the FM-Z and AFM-Z alignments are finite. This is the case of the DW I, AFM I, and AFM II states. In contrast, the DW II, NM, and NPI fail to feature both FM-Z and AFM-Z alignments; their $\expect{Q^{x^2-y^2}_0Q^{x^2-y^2}_1}$ are frozen.

\section{Results for Phantom helix states}\label{sec:results_helix}

\begin{figure}[tbp!]
    \centering
    \includegraphics[width=\linewidth]{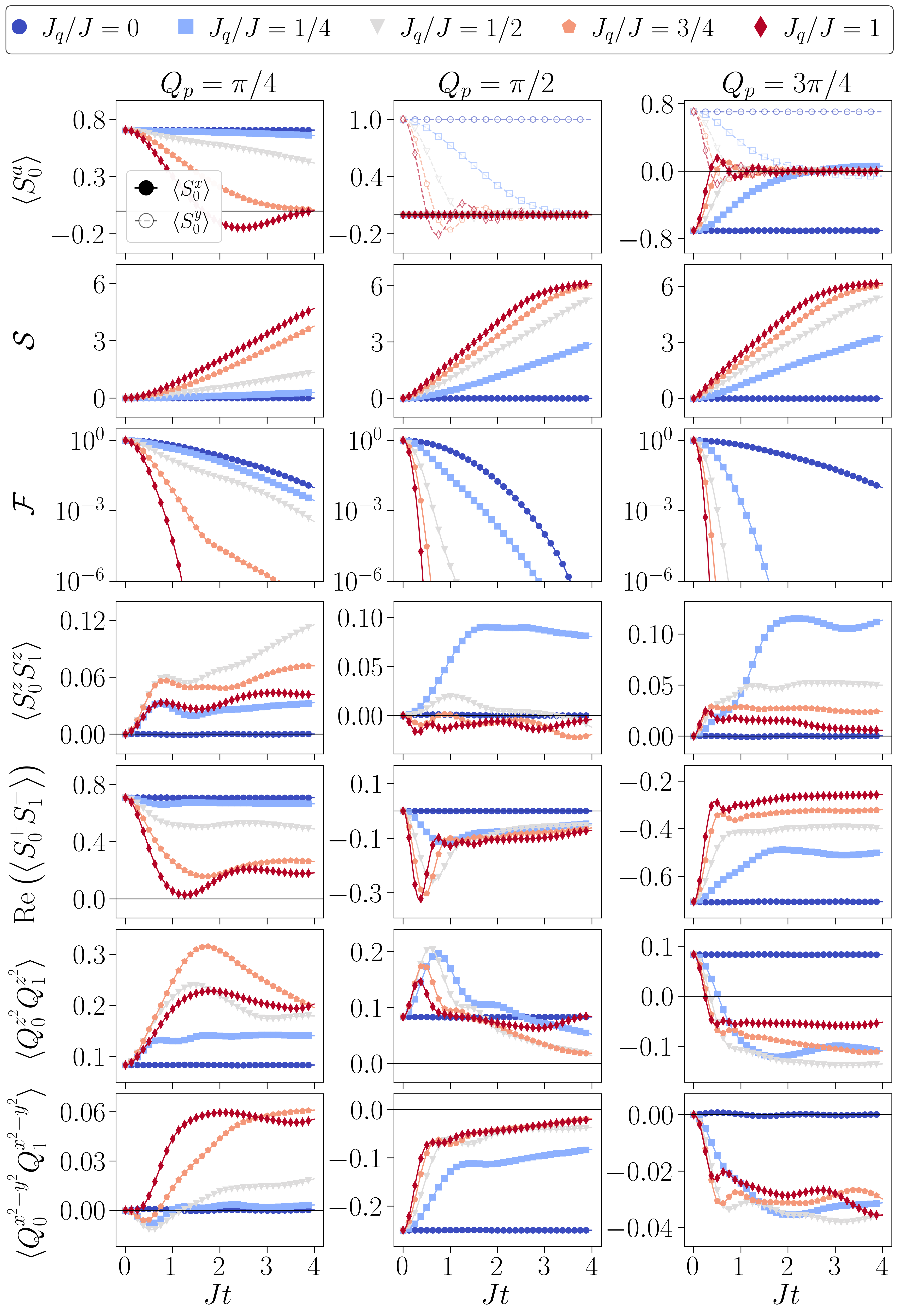}
    \caption{(Color online) Time evolution of observables for phantom helix states with $\theta =\pi/2$ and winding parameters $Q_p=\pi/4$, $Q_p = \pi/2$ and $Q_p=3\pi/4$. (First row) In-plane spin projections $\left\langle S^{x}_{0} \right\rangle$ [solid line with markers] and $\left\langle S^{y}_{0} \right\rangle$ [transparent dashed line], (second row) half-chain entanglement $\mathcal{S}$, (third row) fidelity $\mathcal{F}$, (fourth row) out-of-plane spin correlation $\langle S^{z}_{0}S^{z}_{1}\rangle$, (fifth row) in-plane spin correlation $\text{Re}(\expect{S^{+}_{0}S^{-}_{0}})$, (sixth row) out-of-plane quadrupole correlation $\expect{ Q^{z^2}_{0}Q^{z^2}_{1}}$, and (seventh row) in-plane quadrupole correlation $\expect{Q^{x^2-y^2}_{0}Q^{x^2-y^2}_{1}}$. To satisfy the phantom condition, $J_z/J=\cos(Q_p)$, and we present results for different values of $J_q/J$. Results are presented on lattices of size $L=30$.}
    \label{fig:phantom_physics_correlations}
\end{figure}

In this section, we examine the time-evolution of phantom helix states. Our results are presented in Fig.~\ref{fig:phantom_physics_correlations} for $\theta=\pi/2$ and winding parameters $Q_p = \pi/4$, $\pi/2$, and $3\pi/4$. We now proceed to discuss these results in greater detail. 

We start by noting that since $\expect{S^{z}_{0}} = \cos(\pi/2) = 0$ remains constant throughout the dynamics, the in-plane spin projections, $\expect{S^{x}_{0}}$ and $\expect {S^{y}_{0}}$, better illustrate the system's evolution. These are presented in the first row of Fig.~\ref{fig:phantom_physics_correlations} at the center of the chain. At $Q_p = \pi/4$, we observe that $\expect{S^{x}_{0}}=\expect{S^{y}_{0}}$ at all times. At $Q_p =\pi/2$, we see that although $\expect{S^{y}_{0}}$ exhibits a time-dependent behavior, $\expect{S^{x}_0}$ remains constant at zero. At $Q_p = 3\pi/4$ the evolution of the $x$ and $y$-axis projection differ by a sign, $\expect{S^{x}_{0}}=-\expect{S^{y}_{0}}$.
In the $J_q/J = 0$ limit, the curves freeze at their initial values, consistent with the fact that phantom helix states are \textit{approximate} eigenstates of the Hamiltonian in this limit for OBC. As the quadrupolar terms are introduced, $\expect{S^{x}_{0}}$ and $\expect {S^{y}_{0}}$ decay towards zero, with a decay rate that increases as $J_q/J$ increases. For every winding parameter, the site begins in a pure dipolar state, satisfying $|\langle \mathbf{S}_{0} \rangle|^2 = 1$. When $J_q/J > 0$, the system evolves toward a purely quadrupolar state where the local magnetization vanishes, $|\langle \mathbf{S}_{0} \rangle|^2 = 0$.

The evolution of the entanglement entropy $\mathcal{S}$ between the two halves of the chain is presented in the second row of Fig.~\ref{fig:phantom_physics_correlations}. In the $J_q/J=0$ limit, $\mathcal{S}=0$ throughout the dynamics, indicating that the two halves of the chain remain unentangled. Conversely, for $J_q/J > 0$, $\mathcal{S}=0$ grows in a similar fashion to the entanglement entropy curves observed in the AFM I and AFM II states (see Fig.~\ref{fig:quad_observables}), where the entanglement entropy grows faster as $J_q/J$ increases. Notably, for $Q_p = \pi/2$ and $Q_p = 3\pi/4$, the entropy rapidly approaches and saturates at $\mathcal{S} \approx 6.4$. This value corresponds to the maximum entanglement entropy attainable under our numerical implementation (see Appendix~\ref{app:decimation}). These results show that the quadrupolar terms in the Hamiltonian accelerate the generation of entanglement across the system.

The third row of Fig.~\ref{fig:phantom_physics_correlations} illustrates the time evolution of the fidelity $\mathcal{F}$. In the $J_{q}/J=0$ limit, $\mathcal{F}$ decreases slightly. Contrary to the freezing of other observables in this limit, the fidelity decays because phantom helix states are not an exact eigenstate of the Hamiltonian when considering OBC. As $J_q/J$ increases, the initial state ``melts'' faster. For example, in the $J_q/J= 1$ limit, the fidelity drops below $10^{-6}$ around $Jt \lesssim 1$. This rapid evolution towards orthogonality mirrors the behavior observed for the AFM I and AFM II initial states (see Fig.~\ref{fig:quad_observables}).

The spin-spin correlation functions, $\expect{ S^{z}_{0}S^{z}_{1}}$ and $\text{Re}(\expect{S^{+}_{0}S^{-}_{1}})$, are displayed in the fourth and fifth rows of Fig.~\ref{fig:phantom_physics_correlations}, respectively. In the $J_{q}/J=0$ limit, the correlations remain frozen at their initial values, $\langle S^{z}_{0}S^{z}_{1} \rangle = 0$ and $\text{Re}(\langle S^{+}_{0}S^{-}_{1} \rangle) = \cos(Q_p)$, consistent with the phantom helix being an \textit{approximate} eigenstate. 

Away from this limit, $\expect{ S^{z}_{0}S^{z}_{1}}$ shows a non-monotonic behavior with $J_q/J$ for $Q_p=\pi/4$ and $3 \pi/4$. For these winding parameter values, the introduction of quadrupolar terms in the Hamiltonian induces the creation of weak FM domains, as evidenced in the evolution of $\expect{ S^{z}_{0}S^{z}_{1}}$ to small positive values. Initially, as $J_q/J$ increases, the magnitude of this correlation grows, reaches its largest value at $J_q/J=1/2$ for $Q_p=\pi/4$, and $J_q/J=1/4$ for $Q_p=3\pi/4$. Then the magnitude decreases as $J_q/J$ further increases. On the contrary, for $Q_p = \pi/2$, $\expect{ S^{z}_{0}S^{z}_{1}}$ evolves towards a small negative value for all values of $J_q/J>0$, except for $J_q/J=1/4$ where the correlation evolves towards a positive value. 

The magnitude of the in-plane correlation function $\text{Re}(\expect{S^{+}_{0}S^{-}_{1}})$ exhibits a decay in magnitude that is more pronounced for larger values of $J_q/J$ for $Q_p = \pi/4$ and $3\pi/4$. As the strength of the quadrupolar coupling increases, the original alignment of in-plane spin vectors is lost more rapidly. The behavior for $Q_p=\pi/2$ is different. First, $\text{Re}(\expect{S^{+}_{0}S^{-}_{1}})$ starts at zero. Then, it rapidly evolves towards a pronounced negative value before $Jt \sim 1$. The time scale for this transient is shorter as $J_q/J$ increases. Finally, $\text{Re}(\expect{S^{+}_{0}S^{-}_{1}})$ exhibits a slow evolution towards zero which roughly independent on $J_q/J$.

The evolution of quadrupolar-quadrupolar correlations $\expect{Q^{z^2}_0 Q^{z^2}_1}$ and $\expect{Q^{x^2-y^2}_{0}Q^{x^2-y^2}_{1}}$ is presented in the sixth and seventh rows of Fig.~\ref{fig:phantom_physics_correlations}. The initial value of $\expect{Q^{z^2}_0 Q^{z^2}_1}$ is $1/12$ for all $Q_p$ considered. This is consistent with having dipolar states with their spin vectors lying in the $xy$-plane (see Table~\ref{tab:magnetic_corr_values}). On the other hand, the initial value of $\expect{Q^{x^2-y^2}_{0}Q^{x^2-y^2}_{0}}$ is zero for $Q_p =\pi/4$ and $Q_p = 3\pi/4$, indicating isotropic in-plane spin fluctuations. However, for $Q_p = \pi/2$, the initial value is $-1/4$. This can be shown by noticing that the bond at the center of the chain is the product state $\ket{Y}\otimes \ket{X}$.

As with previous quantities, the $J_q/J=0$ limit exhibits trivial dynamics, with quadrupolar-quadrupolar correlations remaining frozen at their initial values. When $J_q/J>0$, the evolution of the curves is strongly dependent on the winding parameter and displays a non-monotonic behavior with $J_q/J$. For $Q_p = \pi/4$, both correlation functions are positive. However, the magnitude of $\expect{Q^{x^2-y^2}_{0}Q^{x^2-y^2}_{0}}$ is much smaller than $\expect{Q^{z^2}_{0}Q^{z^2}_{0}}$, indicating that the system favors the development of a quadrupolar moment parallel to the out-of-plane axis. When the winding parameter is $Q_p = \pi/2$, $\expect{Q^{z^2}_{0}Q^{z^2}_{0}}$ reaches a positive maximum before $Jt = 1$, after which the curves evolve towards negative values. We observe that for this winding parameter, the maximum of $\expect{Q^{z^2}_{0}Q^{z^2}_{1}}$ occurs at the same time than that of $\text{Re}(\expect{S^{+}_{0}S^{-}_{1}})$ for every $J_q/J>0$. This is accompanied by a decay in the magnitude of $\expect{Q^{x^2-y^2}_{0}Q^{x^2-y^2}_{0}}$.
Finally, for $Q_p = 3\pi/4$, both quadrupolar correlation functions evolve towards a negative value, showing the system's preference for AFQ-dZ alignment.

\section{Conclusions}\label{sec:conclusions}

In this work, we have presented a comprehensive analysis of the non-equilibrium dynamics of an anisotropic spin-1 chain after a quantum quench. The underlying symmetry of the model was tuned from the non-integrable SU(2) Heisenberg model to the integrable SU(3)- symmetric Heisenberg model via the parameter $J_q/J$, which controls the strength of the quadrupolar interactions. We employed TEBD to numerically examine the dynamics of the local magnetization, entanglement entropy, fidelity, and various correlation functions for a wide set of experimentally accessible initial states. We characterized the dependence of these observables on the parameter $J_q/J$ for eigenstates of the $z$-magnetization operator and for generalized phantom helix states.

A central finding of our investigation is the existence of a new conserved quantity ($M^2$) when $J_q/J=1$. In this limit, the dynamics of the different initial states displayed a great dependence on its initial magnetization $M$ and squared magnetization $M^2$. We demonstrated that these two physical quantities determine the number of accessible states the system has, which in turn sets upper bounds for the entanglement entropy of the system at long times. This analysis, paired with a combinatorial argument for the relative frequency of possible dipolar and quadrupolar alignments, allowed us to understand under which conditions certain correlation functions freeze. Notably, both freezing and fidelity revivals occur when the system is initially prepared in nematic states.

We also studied generalized phantom helix states, which displayed little to no dynamics in the $J_q/J=0$ limit, as they are an \textit{approximate} eigenstate of the Hamiltonian in this limit for OBC. Interestingly, increasing $J_q/J$ leads to faster thermalization, despite the integrablity of the model in the $J_q/J=1$ limit. Our results suggest that further generalizations are required to construct phantom helix states for Hamiltonians with higher symmetries. These generalizations might include the use of the Bethe ansatz, an analytical solution tuned for SU($N$)-symmetric one-dimensional lattices~\cite{Sutherland1975}.

Our results provide a route to realize a rich array of non-equilibrium behavior in spin-1 lattice models, which are accessible to different experimental platforms, such as ultracold alkaline-earth-like atoms in optical lattices in the strongly interacting limit of the SU($N$)-FHM at $1/N$-filling~\cite{EIGP_SC_2025}, the Mott limit of the spin-1 Bose Hubbard model~\cite{Batrouni2009,Pixley2017}, and materials hosting $3d$ transition metal ions with a $d^2$ or $d^8$
configuration such as Y$_2$BaNiO$_5$, where recently collective quadrupolar magnetic excitations have been observed~\cite{Nag2022}.

\begin{acknowledgments}
We thank LSCSC-LANMAC, where part of the computational simulations reported in this work were performed using their HPC server and the grants: DGAPA-UNAM-PAPIIT Grants No. IG101826 and No. IN118823, SECIHTI Grant No. LNC-2023-51, and CONAHCYT-CB Grant No. A1-S-30934. L.E.R.S. acknowledges SECIHTI through funding for graduate studies.

\end{acknowledgments}

\appendix

\section{Equivalence of the $\mathrm{SU}(3)$ Heisenberg model to the Bilinear-Biquadratic model}
\label{app:equivalence_BB_SU3}

In this section we demonstrate the equivalence between the $\mathrm{SU}(3)$ Heisenberg model \eqref{eqn:anisotropic_H} (for $J_z = J_{xy} = J_q = 1$) and the Bilinear-Biquadratic model \eqref{eqn:Bilinear_Biquadratic} with $\gamma = \pi/4$.

We introduce the auxiliary operator:
\begin{equation}
    F^{ab}_{i}=\frac{1}{2}\left( S^{a}_{i}S^{b}_{i}+S^{b}_{i}S^{a}_{i} \right),
\end{equation}
where $a,b$ can take on the values $x,y,z$.

As our initial step, we prove that:
\begin{equation}
\label{eqn:AA_hamBBprime}
    H_{BB'}=J\sum_{i}\left(\frac{1}{2}\mathbf{S}_{i}\cdot\mathbf{S}_{i+1}+\sum_{a,b}F^{ab}_{i}F^{ab}_{i+1} \right)
\end{equation}
is equal to the Bilinear-Biquadratic Hamiltonian with $\gamma=\pi/4$:
\begin{equation}
 \label{eqn:AA_hamBB}  H_{BB}=J\sum_{i}\left(\mathbf{S}_{i}\cdot\mathbf{S}_{i+1}+\left(\mathbf{S}_{i}\cdot\mathbf{S}_{i+1}\right)^2 \right).
\end{equation}
To do this, we expand the product $F^{ab}_{i}F^{ab}_{i+1}$ as follows:
{\small
\begin{equation*}
    F^{ab}_{i}F^{ab}_{i+1}=\frac{1}{4}\sum_{a,b}\left( S^{a}_{i}S^{b}_{i}+S^{b}_{i}S^{a}_{i} \right)\left( S^{a}_{i+1}S^{b}_{i+1}+S^{b}_{i+1}S^{a}_{i+1} \right).
\end{equation*}}
On the other hand, we note that:
\begin{equation*}
    \left(\mathbf{S}_{i}\cdot\mathbf{S}_{i+1} \right)^2=\sum_{a,b}S^{a}_{i}S^{b}_{i}S^{a}_{i+1}S^{b}_{i+1},
\end{equation*}
so that the product $F^{ab}_{i}F^{ab}_{i+1}$ can be written using spin commutation relations as:
\begin{multline*}
\sum_{a,b}F^{ab}_{i}F^{ab}_{i+1}=\left(\mathbf{{S}}_{i}\cdot\mathbf{S}_{i+1}\right)^2\\-\frac{1}{4} \sum_{a,b}\left(\left[S^{a}_{i},S^{b}_{i}\right]\left[S^{a}_{i+1},S^{b}_{i+1}\right]\right),
\end{multline*}
which reduces to:
\begin{equation*}
\sum_{a,b}F^{ab}_{i}F^{ab}_{i+1}=\left(\mathbf{{S}}_{i}\cdot\mathbf{S}_{i+1}\right)^2+\frac{1}{2}\mathbf{S}_{i}\cdot\mathbf{S}_{i+1}.
\end{equation*}
Notice that substituting the above expression for $\sum_{a,b}F^{ab}_{i}F^{ab}_{i+1}$ in equation \eqref{eqn:AA_hamBBprime} yields the Bilinear-Biquadratic hamiltonian:
\begin{equation}
    H_{BB'}=H_{BB}.
\end{equation}

We now prove that the hamiltonian $H_{BB'}$ in eq.~\eqref{eqn:AA_hamBBprime} is equivalent to the $\mathrm{SU}(3)$ Heisenberg Hamiltonian [eq.~\eqref{eqn:anisotropic_H}] with $J_{z}=J_{xy}=J_{q}=J$ :
\begin{equation}
\label{eqn:AA_su3_heis}
    H = J\sum_{i}\left( \mathbf{S}_{i}\cdot\mathbf{S}_{i+1}+\mathbf{Q}_{i}\cdot\mathbf{Q}_{i+1} \right).
\end{equation}

We first note that the auxiliary operator $F^{ab}_{i}$ can be written in terms of the quadrupolar operators as:
\begin{equation}
    F^{ab}_{i}=\frac{1}{2}Q^{ab}_{i},
\end{equation}
when $a\neq b$. Thus, notice that:
\begin{equation*}
    \sum_{a\neq b}F^{ab}_{i}F^{ab}_{i+1}=\frac{1}{2}\sum_{a\neq b}Q^{ab}_{i}Q^{ab}_{i+1}.
\end{equation*}

The following identities express the squared spin operators $\left(S^{a}_{i}\right)^2$ in terms of quadrupolar operators:
\begin{subequations}
\label{eqn:AA_spin_squared_identities}
    \begin{align}
        \left(S^{z}_{i}\right)^2&=\frac{1}{\sqrt{3}}Q^{z^2}_{i}+\frac{2}{3},
        \\
         \left(S^{x}_{i}\right)^2&=\frac{1}{2}Q^{x^2-y^2}_{i}-\frac{1}{2\sqrt{3}}Q^{z^2}_{i}+\frac{2}{3},
         \\
          \left(S^{y}_{i}\right)^2&=-\frac{1}{2}Q^{x^2-y^2}_{i}-\frac{1}{2\sqrt{3}}Q^{z^2}_{i}+\frac{2}{3}.
    \end{align}
\end{subequations}
Identities \eqref{eqn:AA_spin_squared_identities} allow us to write:
\begin{equation}
    \sum_{a}F^{aa}_{i}F^{aa}_{i+1}=\frac{1}{2}Q^{z^2}_{i}Q^{z^2}_{i+1}+\frac{1}{2}Q^{x^2-y^2}_{i}Q^{x^2-y^2}_{i+1}+\frac{4}{3} .
\end{equation}
Thus:
\begin{equation}
     \sum_{a,b}F^{ab}_{i}F^{ab}_{i+1}=\frac{1}{2}\mathbf{Q}_{i}\cdot\mathbf{Q}_{i+1}+\frac{4}{3}.
\end{equation}
Substituting $ \sum_{a,b}F^{ab}_{i}F^{ab}_{i+1}$ in eq.~\eqref{eqn:AA_hamBBprime} and identifying the $\mathrm{SU}(3)$ Heisenberg Hamiltonian \eqref{eqn:AA_su3_heis} yields:
\begin{equation}
    H_{BB'}=\frac{1}{2}H + \frac{4}{3},
\end{equation}
that is, the Bilinear-Biquadratic model at $\gamma = \pi/4$ and the $\mathrm{SU}(3)$ Heisenberg Hamiltonian with $J_z = J_{xy}=J_{q}$  are equivalent. The factor $4/3$ introduces a global phase that cancels when computing expectation values. The factor $1/2$ rescales the system's time scale.

\section{Conserved quantities}
\label{app:conserved_quantities}

\subsection{Net magnetization}
\label{app:net_magnetization}

The magnetization operator, defined as
\begin{equation}
\label{app:AA_magnetization}
    M = \sum_{i}S^{z}_{i},
\end{equation}
commutes with the Hamiltonian, $\left[H, \sum S^{z}_{i}\right]=0$, for every value of its parameters $J_z, J_{xy}$ and $J_q$, and thus provides a conservation law valid for all of the systems considered in this work.

To prove this, we first write the commutation relations of $S^{z}_{i}$ with spin $S^{a}_{i}$ and quadrupolar operators:
\begin{align*}
    \left[S^{z}_{i}, S^{z}_{i} \right]&=0, &  \left[ S^{x}_{i}, S^{z}_{i} \right]&= -\, i S^{y}_{i},
    \\
    \left[S^{y}_{i}, S^{z}_{i} \right]&=iS^{x}_{i} , &  \left[ Q^{z^2}_{i}, S^{z}_{i} \right]&= 0,
    \\
    \left[Q^{x^2-y^2}_{i}, S^{z}_{i} \right]&=-2i Q^{xy}_{i} , &  \left[Q^{xy}_{i}, S^{z}_{i} \right]&= 2iQ^{x^2-y^2}_{i} ,
    \\
    \left[Q^{xz}_{i}, S^{z}_{i} \right]&= - iQ^{yz}_{i}, &  \left[Q^{yz}_{i}, S^{z}_{i} \right]&=  iQ^{xz}_{i} .
\end{align*}

Due to the linearity of the commutator and the structure of $M$ and $H$, we only need to show that $M$ commutes with a generic two-site term in the Hamiltonian:
\begin{equation}
\left[ H, M\right] =   \sum_{i, j} \left[ H_{i, i+1}, S^z_j\right]
\end{equation}

Since spin operators commute if they operate on different sites, we only consider the contributions from $j=i$ and $j=i+1$:
$$
\left[ H,M\right] =  \sum_{i} \left( \left[ H_{i, i+1}, S^z_i\right] + \left[ H_{i, i+1}, S^z_{i+1}\right] \right).
$$
Using the relations above, we compute:
\begin{subequations}
\begin{align}
    \left[ S^{z}_{j}S^{z}_{j+1},  \sum_{i}S^{z}_{i} \right]_{-}&=0,
    \\
    \left[ S^{x}_{j}S^{x}_{j+1}+S^{y}_{j}S^{y}_{j+1},  \sum_{i}S^{z}_{i}\right]_{-}&=0,
    \\
    \left[ \mathbf{Q}_{i}\cdot\mathbf{Q}_{i+1} ,  \sum_{i}S^{z}_{i}\right]_{-}&=0,
\end{align}
\end{subequations}
and thus, $M$ commutes with the Hamiltonian \eqref{eqn:anisotropic_H} for every value of its parameters.

\subsection{Quadratic magnetization}
\label{app:quadratic_magnetization}

We show that the quadratic magnetization operator
\begin{equation}
\label{eqn:AB_squared_magnetization}
   M^2 = \sum_{i} \left(S^{z}_{i}\right)^2
\end{equation}
is a constant of motion when $J_q = J_{xy} =1$ in equation \eqref{eqn:anisotropic_H}. The proof is carried out using the Gell-Mann matrices, the usual representation of the $\mathrm{SU}(3)$ algebra.

The $\mathrm{SU}(3)$ algebra is spanned by eight Hermitian and traceless matrices, given by:
{\footnotesize
	\begin{align*}
		\lambda^1 &= \begin{pmatrix}
			0 & 1 & 0
			\\
			1 & 0 & 0
			\\
			0 & 0 & 0
		\end{pmatrix},&
		\lambda^2 &=\begin{pmatrix}
			0 & - \imath & 0
			\\
			\imath & 0 & 0
			\\
			0 & 0 & 0
		\end{pmatrix},&
		\lambda^3 &=\begin{pmatrix}
			1 & 0 & 0
			\\
			0 & -1 & 0
			\\
			0 & 0 & 0
		\end{pmatrix},\\
		\lambda^4 	&= \begin{pmatrix}
			0 & 0 & 1
			\\
			0 & 0 & 0
			\\
			1 & 0 & 0
		\end{pmatrix},
		&
		\lambda^5&=\begin{pmatrix}
			0 & 0 & -i
			\\
			0 & 0 & 0
			\\
			i & 0 & 0
		\end{pmatrix},&
		\lambda^6 &= \begin{pmatrix}
			0 & 0 & 0
			\\
			0 & 0 & 1
			\\
			0 & 1 & 0
		\end{pmatrix},\\
		\lambda^7 &= \begin{pmatrix}
			0 & 0 & 0
			\\
			0 & 0 & -i
			\\
			0 & i & 0
		\end{pmatrix},&
		\lambda^8 &=\frac{1}{\sqrt{3}} \begin{pmatrix}
			1 & 0 & 0
			\\
			0 & 1 & 0
			\\
			0 & 0 & -2
		\end{pmatrix}.
	\end{align*}}
    These matrices follow the commutation and anticommutation relations \cite{Gilmore2012, Guidry2022}:
    \begin{align}
    \label{eqn:AB_commutation_GM}
        \left[ \lambda^{a}, \lambda^{b} \right]_{-}&=2\imath \sum_{c} f^{abc}\lambda^{c},
        \\
        \left[ \lambda^{a}, \lambda^{b} \right]_{+}&=\frac{4}{3}\delta^{ab} + 2\sum_{c} d^{abc}\lambda^{c}
    \end{align}
    where the antisymmetric and symmetric structure constants are given by  $f^{abc} = -\imath\, \text{Tr}\left( \lambda^a \left[ \lambda^b, \lambda^c \right]_{-} \right)/4$ and $d^{abc}=\text{Tr}\left( \lambda^a \left[ \lambda^b, \lambda^c \right]_{+} \right)/4$.

We define the Gell-Mann matrices acting on site $i$ of the chain as $\mathbf{\Lambda}_i =(\lambda^1_i, \lambda^2_i, \lambda^3_i,\dots, \lambda^7_i, \lambda^8_i )$. This allows us to write the single-site spin operators in terms of the Gell-Mann matrices as:
    \begin{subequations}
    \label{eqn:AB_spin_in_terms_GM}
    \begin{align}
        S^{z}_{i}&=\frac{1}{2}\left(\lambda^3_{i} + \sqrt{3}\lambda^8_{i} \right),
        \\
        S^{x}_{i}&=\frac{1}{\sqrt{2} }\left( \lambda^1_{i} + \lambda^6_{i} \right),
        \\
        S^{y}_{i}&=\frac{1}{\sqrt{2} }\left( \lambda^2_{i} + \lambda^7_{i} \right).
    \end{align}
    \end{subequations}
Using these equalities and the anticommutation relations above, we obtain the following expression for $\left(S^{z}_{i}\right)^2$:
\begin{equation}
    \left(S^{z}_{i}\right)^2 = S^{z}_{i} -\frac{2}{\sqrt{3} }\lambda^{8}_{i}+\frac{2}{3},
\end{equation}
and thus:
\begin{equation}
    M^2 = \sum_{i}\left(S_i^{z} -\frac{2}{\sqrt{3} }\lambda^{8}_{i}+\frac{2}{3}\right).
\end{equation}
Since the magnetization $\sum_{i}S^{z}_{i}$ is a constant of motion, it suffices to show that $\sum_i \lambda^{8}_{i}$ commutes with the Hamiltonian to show that \eqref{eqn:AB_squared_magnetization} is a conserved quantity as well.

The commutation relations of $\lambda^{8}$ with the spin operators can be calculated using the identities \eqref{eqn:AB_spin_in_terms_GM} and the commutation relations \eqref{eqn:AB_commutation_GM}. This yields:
\begin{subequations}
\label{eqn:AB_spin_GM_commutation}
\begin{align}
    \left[S^{z}_{i},\lambda^{8}_{i}\right]_{-}&=0,
    \\
    \left[S^{x}_{i},\lambda^{8}_{i}\right]_{-}&=-i \sqrt{\frac{3}{2}}\lambda^7_{i},
    \\
    \left[ S^y, \lambda^8_{i} \right]_{-}&=i\sqrt{\frac{3}{2}}\lambda^6_{i}.
\end{align}
\end{subequations}

Parting from \eqref{eqn:AB_spin_GM_commutation}, the commutator of $\lambda^8$ and every quadrupolar operator $Q^{\alpha\beta}$ can be calculated, which gives as a result:
\begin{subequations}
\label{eqn:AB_quad_GM_commutation}
\begin{align}
    \left[Q^{z^2}_{i},\lambda^{8}_{i}\right]_{-}&=0, \\
    \left[Q^{x^2-y^2}_{i},\lambda^{8}_{i}\right]_{-}&=-i\sqrt{3}\lambda^5_{i},
    \\
    \left[Q^{xy}_{i},\lambda^{8}_{i}\right]_{-}&=i\sqrt{3}\lambda^4_{i} , \\
    \left[Q^{xz}_{i},\lambda^{8}_{i}\right]_{-}&=i\sqrt{\frac{3}{2}}\lambda^7_{i}  ,
    \\
    \left[Q^{yz}_{i},\lambda^{8}_{i}\right]_{-}&=-i\sqrt{\frac{3}{2}}\lambda^6_{i}. 
\end{align}
\end{subequations}

Using the relations above, we compute:
\begin{subequations}
\label{eqn:AB_two_site_comm8}
\begin{align}
    \left[ S^{z}_{i}S^{z}_{i+1},  \sum_{i}\lambda^{8}_{i} \right]_{-}&=0,
    \\
    \left[ S^{x}_{i}S^{x}_{i+1}+S^{y}_{i}S^{y}_{i+1},  \sum_{i}\lambda^{8}_{i} \right]_{-}&=A_{i,i+1},
    \\
    \left[ \mathbf{Q}_{i}\cdot\mathbf{Q}_{i+1} ,  \sum_{i}\lambda^{8}_{i} \right]_{-}&=-A_{i,i+1},
\end{align}
\end{subequations}
where $A_{i,i+1}$ is a two-site operator written in terms of the Gell-Mann matrices as:
{\small
\begin{equation}
  A_{i,i+1}=i\sqrt{3}\left( \lambda^2_{i}\lambda^6_{i+1} + \lambda^6_i \lambda^2_{i+1} - \lambda^1_{i}\lambda^7_{i+1}-\lambda^7_i \lambda^1_{i+1} \right).
\end{equation}}

Equations \eqref{eqn:AB_two_site_comm8} imply that the commutation of $\sum_{i}\lambda^{8}_{i} $ with the Hamiltonian \eqref{eqn:anisotropic_H} with arbitrary parameters $J_z, J_{xy}$ and $J_q$ is given by:
\begin{equation}
\label{eqn:AB_comm_rel}
    \left[ H, \sum_{i}\lambda_{8} \right]_{-}=\left(J_{xy}-J_{q}\right)\sum_{i}A_{i,i+1}.
\end{equation}
When $J_{xy}=J_{q}$, the commutation relation \eqref{eqn:AB_comm_rel} becomes zero: note that this result is independent of the value of $J_z$. 

The added conservation law of the quadratic magnetization operator $\sum_i \left(S^z_i\right)^2$ has important consequences in the dynamics for every initial state. For instance, the NPI initial state must evolve to another state preserving $\sum\langle \sigma_i \rangle= 1$ and $\sum\langle \sigma^2_i \rangle=1$. Given an $ L$-site chain, there are $L$ eigenstates of the magnetization operator satisfying this condition. Thus, the number of possible entangled states is reduced, decreasing the maximum value the entanglement entropy can reach, as can be seen in Fig.~\ref{fig:quad_observables}.

\begin{figure}[h!]
    \centering
    \includegraphics[width=\linewidth]{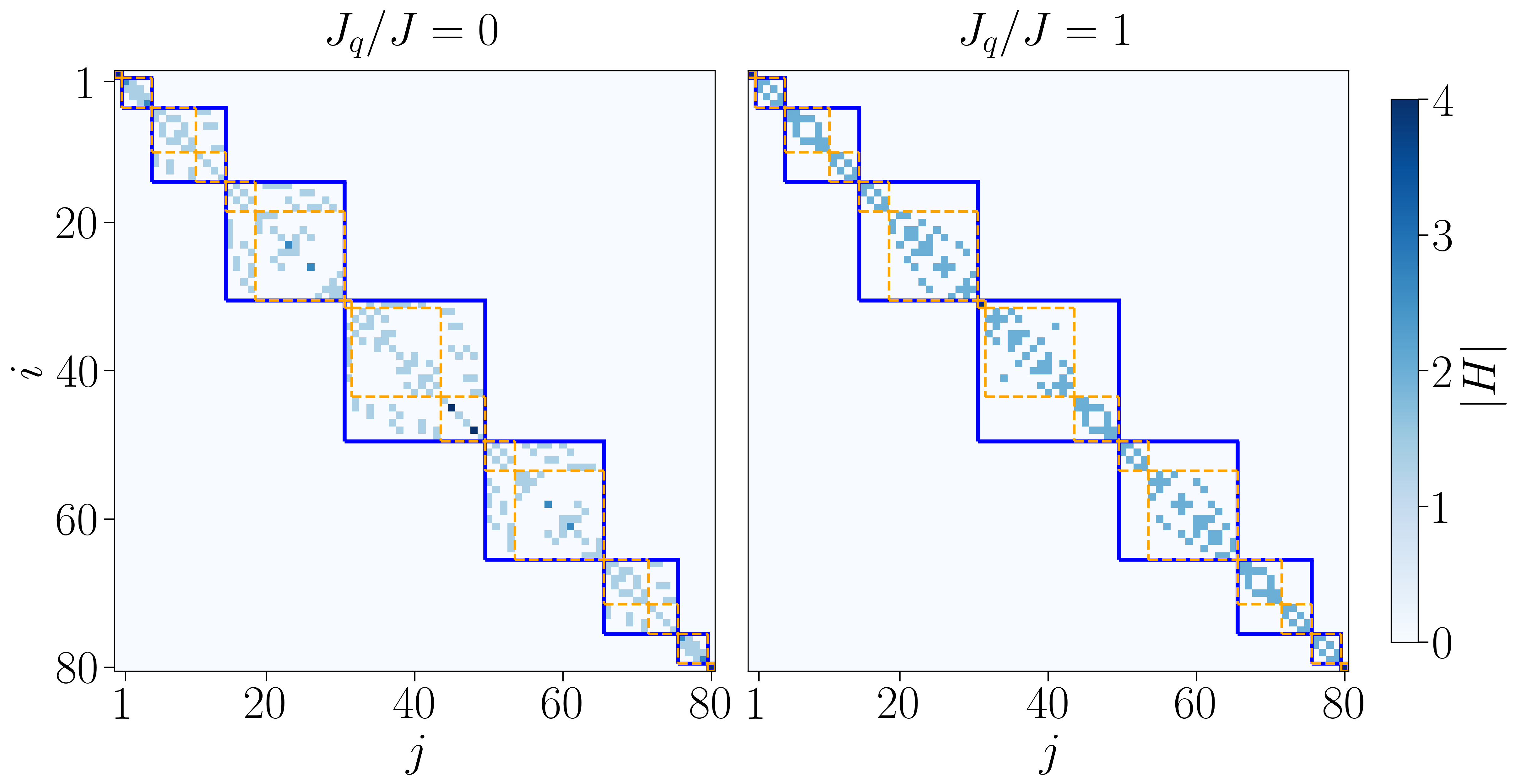}
    \caption{Absolute value of the matrix elements of the Heisenberg Hamiltonian at $J_z=J_{xy}=J$ for $L=4$ and $J_q/J=0,1$. Blue lines indicate magnetization sectors $\sum_i S_i^z$, and orange lines indicate quadratic magnetization sectors $\sum_i (S_i^z)^2$. }
    \label{fig:hamiltonian_subplots}
\end{figure}

Fig.~\ref{fig:hamiltonian_subplots} exemplifies the emergence of the conservation law at $J_q/J = 1$ by showing the plot of the absolute value of the matrix elements for $L=4$. We observe that, in the $J_q/J = 0$ case, the non-zero values lie within magnetization sectors but not within the quadratic magnetization sectors. On the other hand, when $J_q/J=1$, the Hamiltonian is further block diagonalized, i.e. the non-zero values lie within each of the quadratic magnetization sectors.

\section{Finite size errors}
\label{app:finite_size_errors}

\begin{figure}[htbp!]
    \centering
    \includegraphics[width=\linewidth]{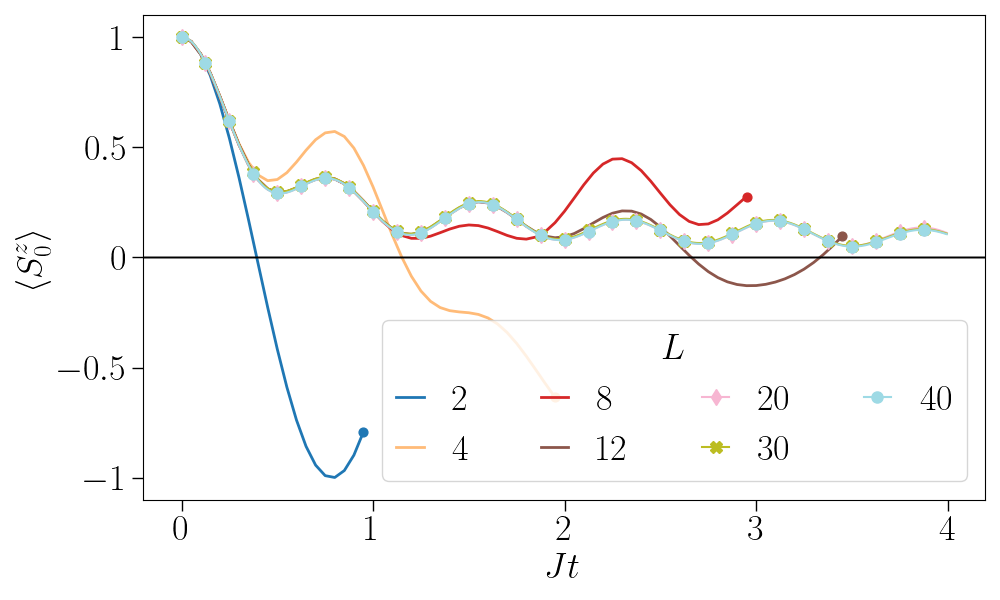}
    \caption{ Time evolution of the out-of-plane spin projection at the middle of the chain $\left\langle S^{z}_{0}\right\rangle$ for the DW I initial state. The chain length ranges from $L=2$ to $L=8$, and a TEBD simulation for $L=20,30$, and $40$ sites, with Trotter step $dt = 0.005$ and $\chi=600$, is included for comparison.}
    \label{fig:finite_size_plot}
\end{figure}

Fig.~\ref{fig:finite_size_plot} shows the results of exact diagonalization for the out-of-plane spin projection at the center of the chain for $L=2,4,8,12$. The TEBD simulation for $L=20,30,40$ and sites is included for comparison. The DW I state was assumed to be the initial state. It is only for short times that the curves coincide: finite-size effects influence the system’s dynamics. Despite restricting the Hilbert space to magnetization sectors and quadratic magnetization sub-blocks, ED requirements scale exponentially with the number of sites $L$.

\section{Details of TEBD calculation}
\label{app:tebd_details}

We now discuss the details of the TEBD algorithm. This section is organized as follows. First, we introduce the MPS representation and the truncation errors associated with its truncation. We then explore the Suzuki-Trotter decomposition and its associated errors. Finally, we provide formulas for computing the expectation values of an operator acting on a single site and on two neighboring sites, as well as a procedure to evaluate the fidelity and the entanglement entropy.

\subsection{Matrix Product States}

Consider a chain of $L$ sites, each with a local Hilbert space dimension $d$. A general quantum state can be written as:
\begin{equation}
    \ket{\psi}= \sum_{\boldsymbol{\sigma}}c^{\boldsymbol{\sigma}}\ket{\boldsymbol{\sigma}}
\end{equation}
where $\boldsymbol{\sigma}=(\sigma_1,\dots,\sigma_i,\dots,\sigma_L)$ denotes the basis configurations and $c^{\boldsymbol{\sigma}}$ are the expansion coefficients $\langle\psi\ket{\boldsymbol{\sigma}}$. A full description of a general spin chain state requires $d^{L}$ coefficients to be accurately described, which grows exponentially with system size $L$. The MPS representation overcomes this difficulty by allowing an efficient state compression \cite{Verstraete2008}.

The MPS representation is achieved by writing the state $\ket\psi$ as a product of matrices:
\begin{equation}
\label{eqn:app_MPS_Form}
\ket{\psi}=\sum_{\boldsymbol{\sigma},\boldsymbol{\alpha}}B^{\sigma_1}_{\alpha_0 \alpha_1 }B^{\sigma_2}_{\alpha_1 \alpha_2} \dots B^{\sigma_L}_{\alpha_{L-1}\alpha_{L}}\ket{\boldsymbol{\sigma}} 
\end{equation}
where $\boldsymbol{\alpha}=(\alpha_0,\alpha_1,\alpha_2,\dots,\alpha_L)$ are the bond indices of the rank-3 tensors $B^{\sigma_i}_{\alpha_{i-1}\alpha_{i}}$. In eq.~\eqref{eqn:app_MPS_Form} we assume a right-canonical normalization of the $B^{\sigma_i}_{\alpha_{i-1}\alpha_{i}}$ tensors, given by:
\begin{equation}
    \sum_{\sigma_i \alpha_i} B^{\sigma_i}_{\alpha_{i-1}\alpha_{i} }B^{\dagger\,\sigma_{i}}_{\alpha_i\alpha_{i-1}}=I_{\alpha_{i-1}\alpha_{i-1}},
\end{equation}
where $B^{\dagger\,\sigma_{i}}_{\alpha_i\alpha_{i-1}}$ is the complex conjugate of $ B^{\sigma_i}_{\alpha_{i-1}\alpha_{i} }$.

We control a parameter $\chi$, known as \textit{maximum bond dimension}. The maximum entanglement entropy is dependent on $\chi$. This parameter sets an upper bound to the $\boldsymbol{\alpha}$ bond indices: if any matrix index exceeds this set value, it is truncated. The numerical errors due to the truncation of the MPS in the context of our simulations are discussed in Appendix \ref{app:decimation}.

The tensors $B^{\sigma_i}$ are constructed via iterative singular value decomposition (SVD) of the state coefficients $c^{\boldsymbol{\sigma}}$. As a first step, these coefficients are reshaped into a matrix of dimension $d^{L-1}\times d$:
\begin{equation*}
    \psi^{(\sigma_{1}\dots\sigma_{L-1}),\sigma_{L}}=\text{reshape}(c^{\sigma_{1}\dots\sigma_{L-1}\sigma_{L}}).
\end{equation*}
The SVD of matrix $\psi$ yields:
\begin{equation*}
    \psi^{(\sigma_{1}\dots\sigma_{L-1}),\sigma_{L}}=\sum_{\alpha_{L-1} }U^{\sigma_1 \dots \sigma_{L-1} }_{\alpha_{L-1}}\Lambda^{L}_{\alpha_{L-1}\alpha_{L-1}}\left(V^{\sigma_L}_{\alpha_{L-1}\alpha_{L}}\right)^{\dagger}.
\end{equation*}

The first and rightmost MPS tensor is defined as $B^{\sigma_L}_{\alpha_{L-1}\alpha_{L}}=\left(V^{\sigma_L}_{\alpha_{L-1}\alpha_{L}}\right)^{\dagger}$. For the next step, we define $\psi^{(\sigma_{1},\sigma_{2}\dots\sigma_{L-1})}=U^{(\sigma_1 \dots \sigma_{L-1} )}_{\alpha_{L-1}}\Lambda^{L}_{\alpha_{L-1}}$. This new matrix is then reshaped and decomposed via SVD to obtain $B^{\sigma_{L-1}}_{\alpha_{L-2}\alpha_{L-1}}$ and $\Lambda^{L-1}_{\alpha_{L-2}\alpha_{L-2}}$. This process is repeated until the full state is expressed as in equation \eqref{eqn:app_MPS_Form}.

In every step of the process, it is important to save the diagonal matrices $\Lambda^{i}$. Its squared elements $\lambda^2_{j}$ correspond to the eigenvalues for the bipartition at bond $i$. Thus, these singular values allow the computation of the entanglement entropy across the chain.

\subsection{Suzuki-Trotter decomposition}

The time evolution of a quantum state $\ket{\psi(t)}$ is governed by the unitary operator $U(dt)=e^{-iHdt}$, such that $\ket{\psi(t+dt)}=U(dt)\ket{\psi(t)}$. We consider the action of this operator on a quantum state in MPS form for an infinitesimal time step $dt$.

Spin chain Hamiltonians with nearest-neighbor interactions can be written as $H = H_{\text{even}}+H_{\text{odd}}$, where $H_{\text{even}}$ and $H_{\text{odd}}$ contain the even and odd terms, respectively. That is, letting $h_{i,i+1}$ represent the interaction between two adjacent sites, then:
\begin{equation}
\label{eqn:AP_hamilton_evn_odd}
H_{\text{even}} = \sum_{i \text{ even}}h_{i,i+1},\quad H_{\text{odd}}= \sum_{i \text{ odd}}h_{i,i+1}.
\end{equation}
Note that $[h_{i,i+1},h_{j,j+1}]$ commute if both $i,j$ are even (or odd). Thus, all of the terms in $H_{\text{even}}$ (and $H_{\text{odd}}$) commute with each other. However, $H_{\text{even}}$ and $H_{\text{odd}}$ do not commute in general, i.e., $[H_{\text{even}},H_{\text{odd}}]\neq 0$.

Using the first-order Suzuki-Trotter decomposition, the time evolution operator is approximated as:
\begin{equation}
\label{eqn:AA_suzuki_trotter}
    e^{-iHdt}= e^{-iH_{\text{even}}dt}e^{-iH_{\text{odd}}dt}+\mathcal{O}(dt^2).
\end{equation}
Since every term in $H_{\text{even}}$ and $H_{\text{odd}}$ commute, the exponential operators can be factorized exactly:
\begin{align*}
    e^{-iH_{\text{even}}dt}&=\prod_{i\,\text{even}}e^{-ih_{i,i+1}dt},\\
     e^{-iH_{\text{odd}}dt}&=\prod_{i\,\text{odd}}e^{-ih_{i,i+1}dt}.
\end{align*}
Substituting into eq.~\eqref{eqn:AA_suzuki_trotter} yields the total evolution operator $U(dt)$ taking the form of a sequence of local operators:
\begin{equation}
\label{eqn:trotter_decomp}
U(dt) \approx  \left[\prod_{n \text{ odd}}U_{i,i+1}(dt) \right]\left[\prod_{i \,\text{even}}U_{i,i+1}(dt) \right],
\end{equation}
where:
\begin{equation}
\label{eqn:u_dos_sitios}
U_{i,i+1}=e^{-ih_{i,i+1}dt}.
\end{equation}
Eq.~\eqref{eqn:trotter_decomp} reduces the global time evolution to a series of operations applied bond by bond.

To evaluate the action of $U_{i,i+1}$ on a bond $i$, we first contract neighboring MPS tensors $B^{\sigma_{i}}_{\alpha_{i}\alpha_{i+1}}$ and $B^{\sigma_{i+1}}_{\alpha_{i+1}\alpha_{i+2}}$ with the singular value matrix $\Lambda^{i}$ to form the 4-index tensor $\Theta$:
\begin{equation}
    \Theta^{\sigma_{i}\sigma_{i+1}}_{\alpha_{i-1}\alpha_{i+1}}=\sum_{\alpha_{i}\tilde{\alpha}_{i-1}}\Lambda^{i}_{\alpha_{i-1}\tilde{\alpha}_{i-1}}B^{\sigma_{i}}_{\tilde\alpha_{i-1}\alpha_{i}}B^{\sigma_{i+1}}_{\alpha_{i}\alpha_{i+1}}.
\end{equation}

The time evolution is then applied by contracting $\Theta$ with the single-bond time evolution operator $U_{i,i+1}$ reshaped as a 4-index tensor:
\begin{equation*}
    \tilde\Theta^{\sigma_i\sigma_{i+1}}_{\alpha_{i-1}\alpha_{i+1}}=\sum_{\tilde\sigma_i \tilde\sigma_{i+1}}U^{\sigma_i \sigma_{i+1}}_{\tilde\sigma_{i}\tilde\sigma_{i+1}}\Theta^{\tilde\sigma_{i}\tilde\sigma_{i+1}}_{\alpha_{i-1}\alpha_{i+1}}
\end{equation*}

To restore the MPS form, we perform a SVD on $\tilde \Theta$. By reshaping the tensor into a matrix and decomposing it, we obtain:
{
\begin{equation}
    \tilde\Theta^{\sigma_i, \sigma_{i+1}}_{\alpha_{i-1},\alpha_{i+1}}=\sum_{\tilde\alpha}A^{\sigma_i}_{\alpha_{i-1},\tilde\alpha}\tilde{\Lambda}^{i+1}_{\tilde\alpha,\tilde\alpha}\left(V^{\sigma_{i+1}}_{\tilde\alpha,\alpha_{i+1}} \right)^{\dagger}.
\end{equation}}
This gives the updated local tensors at bond $i$:
\begin{subequations}
    \begin{align}
        \tilde{B}^{\sigma_{i}}_{\alpha_{i-1}\alpha_{i}}&=\left(\Lambda^{i} \right)^{-1}_{\alpha_{i-1}\alpha_{i-1}}A_{\alpha_{i-1}\alpha_{i}}^{\sigma_i}\tilde\Lambda^{i+1}_{\alpha_{i}\alpha_{i}},
        \\
        \tilde B^{\sigma_{i+1}}_{\alpha_{i}\alpha_{i+1}}&=\left(V_{\alpha_{i}\alpha_{i+1}}^{\sigma_{i+1}} \right)^{\dagger}.
    \end{align}
\end{subequations}
By iterating this local update procedure across all odd and even bonds following the sequence defined in eq.~\eqref{eqn:trotter_decomp}, the full quantum state is propagated forward by a single time step $dt$.

\subsection{Calculation of observables and physical measures}
\label{app:tebd_observables}

The evolution of relevant observables and physical measures can be computed using the updated MPS representation. The procedure depends on whether the operator associated with this observable acts on a single site or on two sites. Other measures rely on the inner product between two MPS (e.g., the fidelity) or on the singular values obtained from an SVD (e.g., the entanglement entropy). This section shows how the expected values of the single-site and two-neighboring-site operators are computed, along with the fidelity and entanglement entropy.

\subsubsection{Expected value of single-site and neighboring operators}

To calculate the expected value of single site operators\footnote{The spin projection operators $S^{z}_{i}$, $S^{x}_{i}$ and $S^{y}_{i}$ are examples of single-site operators.}we define the three-index tensor $\Phi$:
\begin{equation}
    \Phi^{\sigma_{i}}_{\alpha_{i-1}\alpha_{i+1}}=\sum_{\alpha_{i}}\Lambda^{i}_{\alpha_{i-1}\alpha_{i-1}}B^{\sigma_i}_{\alpha_{i-1}\alpha_{i}}.
\end{equation}
Assuming that the matrix representation of the single-site operator is given by $\mathcal{O}^{\sigma_i\sigma_i}$, the expected value is given by:
\begin{equation}
	\langle \mathcal{O}\rangle=\sum_{\sigma_{i}\tilde\sigma_{i},\alpha_{i-1},\alpha_{i+1}}\left(\Phi^{\sigma_{i}}_{\alpha_{i-1}\alpha_{i+1}}\right)^{*}\mathcal{O}^{\sigma_i\tilde\sigma_i}\Phi^{\tilde\sigma_i}_{\alpha_{i-1}\alpha_{i+1}.}
\end{equation}

We now assume an operator acting on two neighboring sites~\footnote{For example, the correlation functions $ S^{z}_{i}S^{z}_{i+1}$, $S^{+}_{i}S^{+}_{i+1}$, $Q^{z^2}_{i}Q^{z^2}_{i+1}$, and $Q^{x^2-y^2}_{i}Q^{x^2-y^2}_{i+1}$.} with $\mathcal{O}^{\tilde{\sigma}_i \tilde\sigma_{i+1}}_{\sigma_i \sigma_{i+1}}$ its matrix representation. The expected value takes the following form:
\begin{equation}
	\langle \mathcal{O}\rangle=\sum^{\sigma_{i+1}\sigma_i\tilde\sigma_i\tilde\sigma_{i+1}}_{{\alpha_{i}\alpha_{i+2}\tilde\alpha_i\tilde\alpha_{i+2}}}\left(\Xi^{\sigma_{i+1}\sigma_{i}}_{\alpha_{i+2}\tilde \alpha_{i}} \right)^{\dagger}C_{\tilde\alpha_{i}\alpha_{i}}\Xi^{\sigma_i\sigma_{i+1}}_{\alpha_{i}\alpha_{i+2}}\mathcal{O}^{\tilde\sigma_i \tilde\sigma_{i+1}}_{ \sigma_i \sigma_{i+1}}
\end{equation}
where:
\begin{equation}
    \Xi^{\sigma_i\sigma_{i+1}}_{\alpha_{i}\alpha_{i+2}}=\sum_{\alpha_{i+1}}B^{\sigma_i}_{\alpha_i\alpha_{i+1}}B^{\sigma_{i+1}}_{\alpha_{i+1}\alpha_{i+2}},
\end{equation}
is the contraction of the $B$ matrices at the two neighboring sites, and $C_{\tilde\alpha_{i}\alpha_{i}}$ is defined recursively as:
\begin{equation}
	C_{\tilde\alpha_{i}\alpha_{i}}=\sum_{\sigma_i \tilde\alpha_i\alpha_{i}}B^{\dagger\,\sigma_i}_{\tilde\alpha_{i+1}\tilde\alpha_{i}}C_{\tilde \alpha_{i}\alpha_{i}}B^{\sigma_i}_{\alpha_i \alpha_{i+1}},
\end{equation}

\subsubsection{Fidelity calculation}

To calculate the fidelity at a time $t$, the inner product of the initial state and the state at time $t$ must be computed. We assume both of these states are written in their MPS representation:
\begin{subequations}
    \begin{align}
        \ket{\psi(0)}&=\sum_{\boldsymbol{\alpha}}B^{\sigma_{1}}_{\alpha_0 \alpha_1}\dots B^{\sigma_{L}}_{\alpha_{L-1}\alpha_{L}}\ket{\boldsymbol{\sigma}},
        \\
        \ket{\psi(t)}&=\sum_{\boldsymbol{\beta}}\tilde B^{\sigma_{1}}_{\beta_0 \beta_1}\dots \tilde B^{\sigma_{L}}_{\beta_{L-1}\beta_{L}}\ket{\boldsymbol{\sigma}}.
    \end{align}
\end{subequations}
We define an auxiliary tensor given by:
\begin{equation}
    E^{[2]}_{\alpha_2\beta_2}=\sum_{\sigma_1 \gamma}\left( B^{\sigma_1}_{\alpha_2 \gamma} \right)^{\dagger}\tilde B^{\sigma_1}_{\gamma\beta_2},
\end{equation}
where the contraction over $\gamma$ is valid since the $\alpha_0$ and $\beta_0$ bond dimensions are always 1.

We now calculate the tensor $E^{[i]}_{\beta_i\alpha_i}$ recursively:
\begin{equation}
    E^{[i]}_{\alpha_i\beta_i} = \sum_{\sigma_{i}\alpha_{i-1}\beta_{i-1}}\left(B^{\sigma_i}_{\alpha_{i}\alpha_{i-1}} \right)^{\dagger}\tilde B^{\sigma_i}_{\beta_{i-1}\beta_{i}}E^{[i-1]}_{\alpha_{i-1}\beta_{i-1}}.
\end{equation}
The inner product is given by:
\begin{equation}
    \bra{\psi(0)}\psi(t)\rangle = E^{[L+1]},
\end{equation}
and so the fidelity is calculated as $\mathcal{F}=\left|\bra{\psi(0)}\psi(t)\rangle\right|^2$.

\subsection{Truncation errors}
\label{app:decimation}

An approximation error in the TEBD algorithm occurs when the MPS matrices are truncated because their bond dimension exceeds the maximum bond allowed $\chi$. This error increases as the entropy approaches the numerical limit set by $S_{\max} = \log(\chi)$.

\begin{figure*}[htbp!]
    \centering
    \includegraphics[width=0.9\linewidth]{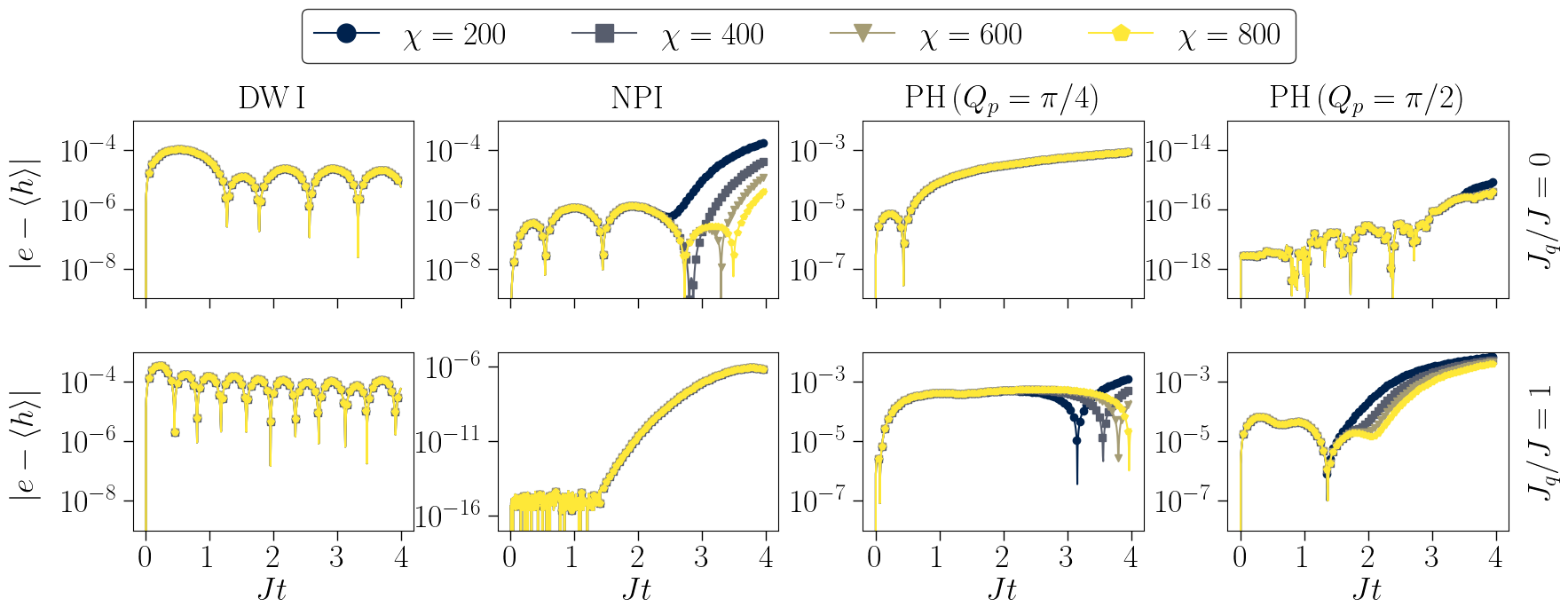}
    \caption{ Absolute difference of the initial energy by site $e$ for the given initial state and its numerical evolution $\langle h\rangle$ at time $t$. The maximum bond dimensions considered were $\chi=200$, $\chi = 600$ and $\chi = 800$, with a Trotter time step $dt = 5\times 10^{-3}$. The initial states considered are the DW I and NPI magnetic states, as well as the $Q_p = \pi/4,\pi/2$ phantom helix states. Results are presented for $J_{q}/J=0$ and $J_q/J = 1$ ($J=J_z=J_{xy}$ for DW I, NPI, and $J=J_{xy}$ for phantom helix states). } 
    \label{fig:chi_error_energy}
\end{figure*}

\begin{figure*}[htbp!]
    \centering
    \includegraphics[width=0.9\linewidth]{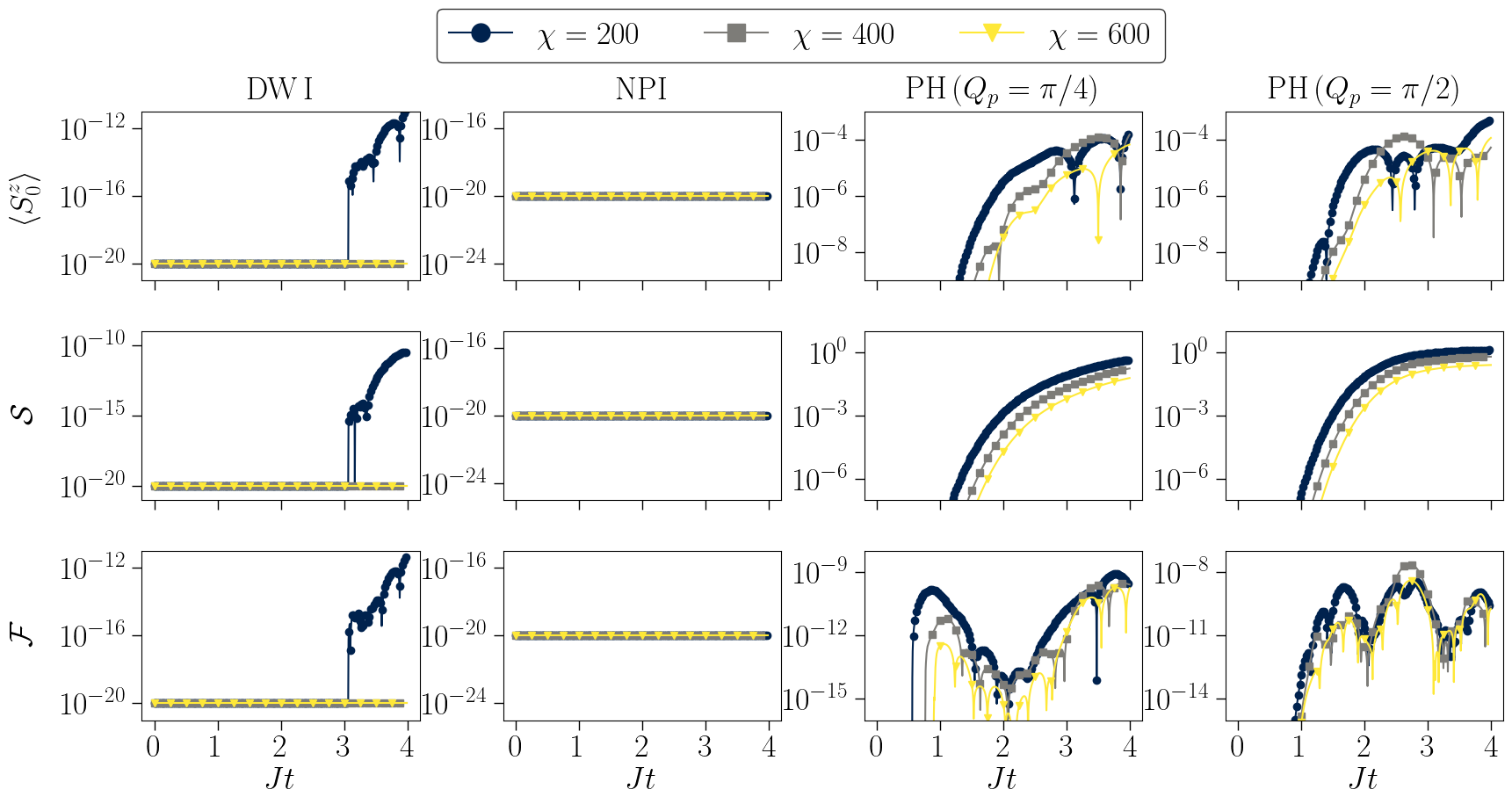}
    \caption{Absolute difference in local magnetization $\expect{S^z_0}$, entanglement entropy $\mathcal{S}$, and fidelity $\mathcal{F}$ for TEBD simulations of spin chains in three initial states (DW I, NPI, and phantom with $Q_p=\pi/4$ and $\pi/2$). Different maximum bond dimensions (\(\chi =200,600,400\)) are compared with respect to a reference simulation with a bigger bond dimension (\(\chi= 800\). In all simulations, $J_{q}/J_{xy}=1$. For the DW I and NPI states, $J=J_z=J_{xy}$, while for the phantom helix state $J_z/J_{xy} = \cos(Q_p)=0$ and $J=J_{xy}$. A small value of $\varepsilon=10^{-20}$ was added to all differences.} 
    \label{fig:chi_error_observable}
\end{figure*}

Time evolution under a time-independent Hamiltonian should preserve the system's initial energy. Therefore, the evolution of the energy in our system serves as a means to assess the convergence of the values of $dt$ and $\chi$ employed in our work. Fig.~\ref{fig:chi_error_energy} illustrates the difference between the energy by site of the initial state $e=\bra{\psi(0)}H|\ket{\psi(0)}/L $ and the numerically calculated energy by site at a later time $\langle h\rangle = \bra{\psi(t)}H|\ket{\psi(t)}/L $. The Trotter time step $dt$ is set to $5\times 10^{-3}$, and the maximum bond dimensions are $200$, $400$, $600$ and $800$.

The results in Fig.~\ref{fig:chi_error_energy} show that the Trotter error associated to the time step $dt = 5\times 10^{-3}$ dominates at small times $Jt <3$. An interesting case is the DW I state (first column), where the absolute difference evolution curve is exactly the same for every value of $\chi$. The NPI initial state (second column) shows the same behavior for $J_q/J = 1$, while $J_q/J=0$ curves show that the decimation errors are more pronounced at $Jt>2$ for lower values of $\chi$. Also note that phantom helix states (third and fourth column) show more pronounced decimation errors for $J_q/J = 1$.

The convergence of the system's time evolution can also be assesed by looking a the fluctuations in local observables with respect to a simulation with a higher value of $\chi$. We show in Fig.~\ref{fig:chi_error_observable} the difference in the value for various physical observables between the simulations with $\chi$ and the largest maximum bond dimension, $\chi=800$.

Fig.~\ref{fig:chi_error_observable} shows that increasing the value of $\chi$ has little to no effect in the DW I (first column) and NPI state (second column), thus proving that the Trotter error in these simulations dominates. Since the entanglement entropy $\mathcal{S}$ (see Fig. \ref{fig:quad_observables}) in both of these magnetic states grows slowly in the $J_q/J=1$ limit, decimation errors do not play an important role. The opposite is the case for the phantom helix states (third and fourth columns) where the error for the local magnetization grows to an order of $10^{-3}$. Note that this error is more pronounced for lower values of $\chi$, thus showing that states with high entanglement are more prone to decimation errors.

\subsection{Trotter approximation errors}
\label{app:trotter_explanation}

In contrast to truncation errors, the Trotter error manifests during the short-time evolution. 
Scaling as $\mathcal{O}(dt^2)$ [see Eq.~\eqref{eqn:AA_suzuki_trotter}], this error implies a strict trade-off: 
while reducing the time step $dt$ enhances numerical precision, it significantly increases the computational cost.

\begin{figure*}[htbp!]
    \centering
    \includegraphics[width=0.9\linewidth]{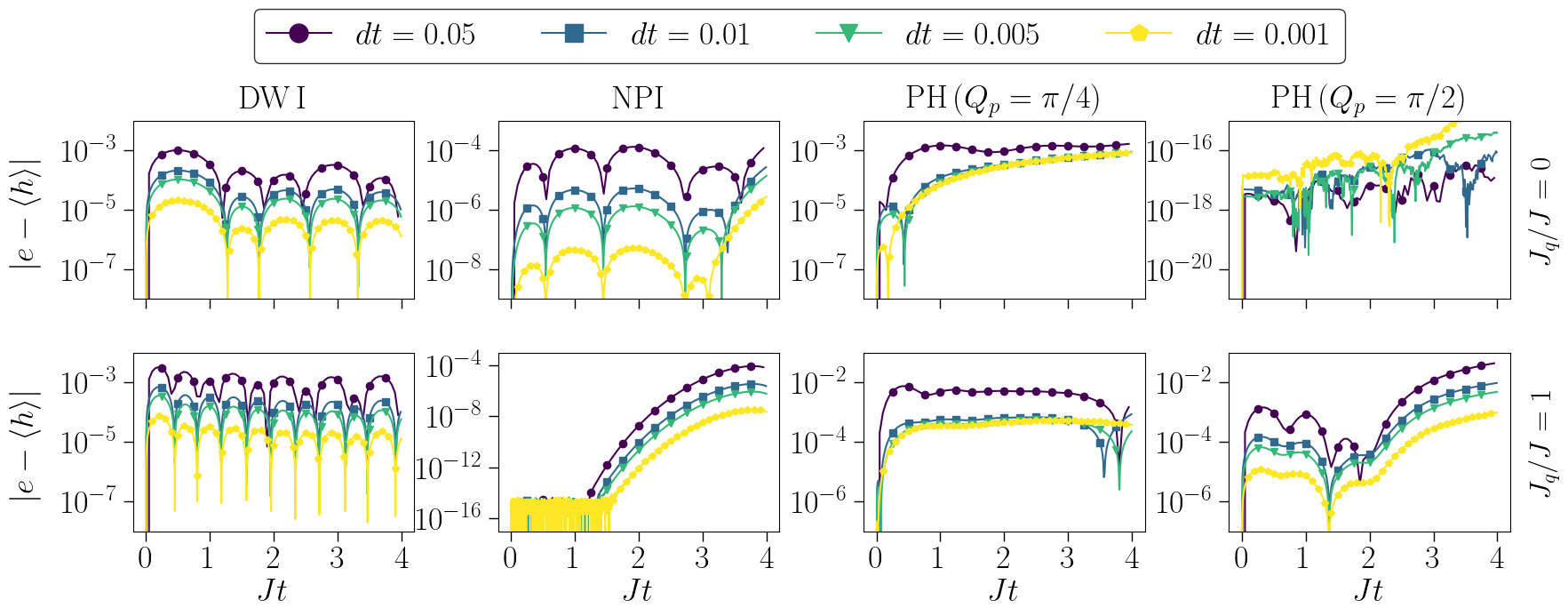}
    \caption{Absolute difference of the initial energy by site $e$ for the given initial state and its numerical evolution $\langle h\rangle$ at time $t$. The Trotter time steps considered were $dt=0.05$, $dt = 0.01$, $dt = 0.005$ and $dt = 0.001$. The initial states considered are the DW and OI magnetic states, as well as the $Q_p = \pi/4,\pi/2$ phantom helix states. The $\mathrm{SU}(2)$ Heisenberg model $J_q = 0$ and its  $\mathrm{SU}(2)$ limit $J_q = 1$ are considered. } 
    \label{fig:trotter_error_energy}
\end{figure*}

\begin{figure*}[htbp!]
    \centering
    \includegraphics[width=0.9\linewidth]{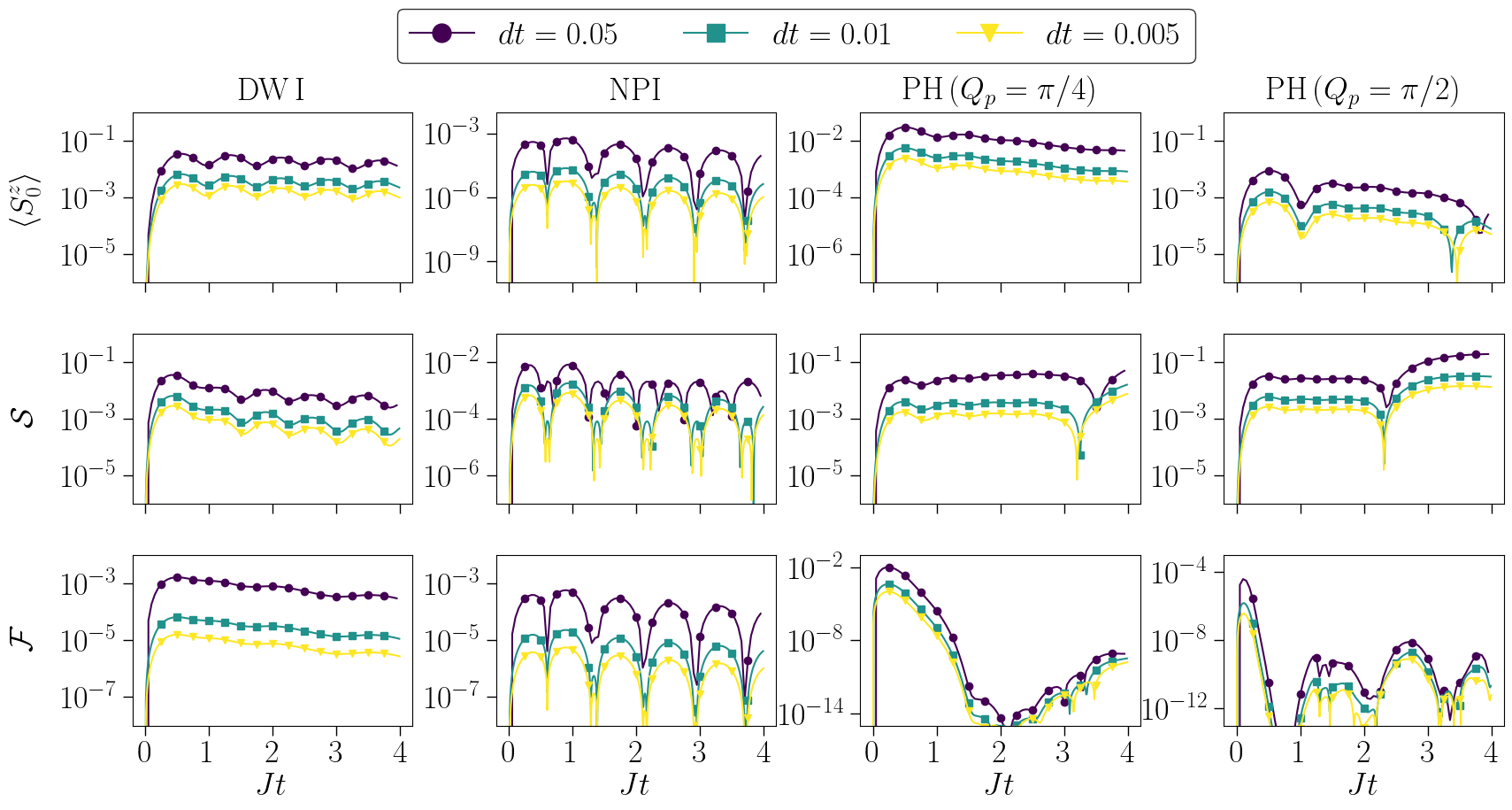}
    \caption{Absolute difference in local magnetization $\expect{S^z_0}$, entanglement entropy $\mathcal{S}$, and fidelity $\mathcal{F}$ for TEBD simulations of spin chains in four initial states (DW I, NPI, and phantom with $Q_p=\pi/4$ and $\pi/2$). Different time steps (\(dt = 5\times 10^{-2}, 10^{-2}, 5\times 10^{-3}\)) are compared with respect to a reference simulation with a finer time step (\(dt = 10^{-3}\). In all simulations, $J_{xy}=J_{q}$. For the DW I and NPI states, $J_z=J_{xy}$, while for the phantom helix state $J_z/J_{xy} = \cos(Q_p)=0$.} 
    \label{fig:trotter_error_observable}
\end{figure*}

Fig.~\ref{fig:trotter_error_energy} shows the absolute difference between the initial energy by site $e $ and the numerically calculated energy by site at a later time $\langle h\rangle$. Trotter time steps $dt = 5 \times 10^{-2}$, $10^{-2}$, $5\times 10^{-3}$ and $10^{-3}$ are considered, with $\chi = 600$ fixed.

The difference between the initial energy $e$ and its numerically computed evolution $\langle h\rangle$ reaches maximum values of approximately $10^{3}$ for magnetic initial states and around $10^{-2}$ for phantom helix states, using the biggest Trotter step, $dt = 5\times 10^{-2}$. Reducing the Trotter step to $dt = 5\times 10^{-3}$ decreases the difference by up to one order of magnitude, and to $dt=10^{-3}$ by up to four orders of magnitude. The difference for $dt = 10^{-2}$ lies between those of $dt = 5\times 10^{-2}$ and $dt = 5\times 10^{-3}$. Moreover, in the case of the phantom helix with $Q_p = \pi/2$ and $J_q = 0$, the error is smaller than the typical floating-point tolerance value, $\epsilon = 10^{-12}$.

Fig.~\ref{fig:trotter_error_observable} shows the difference between the numerical results for the out-of-plane spin projection $\langle S^{z}_{0}\rangle$, the entanglement entropy $\mathcal{S}$ and the fidelity with time steps $dt=5\times 10^{-2},10^{-2},5\times 10^{-3}$ against a reference simulation with $dt=10^{-3}$. Note that the fluctuations with respect to the finer time step are more pronounced for bigger values of $dt$. The evolution of these discrepancies is swift and reach its maximum values for $Jt <1$.

\bibliography{Spin_chain}

 \end{document}